\documentstyle[11pt,amsmath,amssymb,epsf,epic]{article}

\numberwithin{equation}{section}

\newcommand{\be}{\begin{equation}}
\newcommand{\ee}{\end{equation}}
\newcommand{\bea}{\begin{eqnarray}}
\newcommand{\eea}{\end{eqnarray}}

\newcommand{\e}{{\rm e}}

\renewcommand{\d}{{\rm d}}

\newcommand{\grintl}{[\kern-.18em [}
\newcommand{\grintr}{]\kern-.18em ]}

\newcounter{resultcounter}[section]

\newtheorem{thm}[resultcounter]{Theorem}
\newtheorem{lem}[resultcounter]{Lemma}
\newtheorem{prop}[resultcounter]{Proposition}

\newtheorem{definition}[resultcounter]{Definition}

\def\bed{\begin{definition}}
\def\eed{\end{definition}}

 \def\cH{{\cal H}}

\newcommand{\r}{{\rm R}}
\newcommand{\s}{{\rm S}}
\newcommand{\h}{{\cal H}}
\newcommand{\cx}{{\mathbb C}}
\newcommand{\rx}{{\mathbb R}}
\renewcommand{\i}{{\rm i}}

\newcommand{\fer}[1]{(\ref{#1})}
\newcommand{\scalprod}[2]{\left\langle {#1}, {#2}\right\rangle}
\newcommand{\bbbone}{\mathchoice {\rm 1\mskip-4mu l} {\rm 1\mskip-4mu l}
{\rm 1\mskip-4.5mu l} {\rm 1\mskip-5mu l}}
\newcommand{\mm}{{\mathfrak M}}

\begin{document}
\title{Entanglement Evolution via\\
 Quantum Resonances}
\author{
Marco Merkli\footnote{Email: merkli@mun.ca, webpage:
http://www.math.mun.ca/\ $\widetilde{}$\,merkli/}\,
\footnote{Supported by NSERC
Discovery Grant 205247}\\
Department of Mathematics and Statistics\\
 Memorial University \\
 St. John's, Newfoundland, A1C 5S7 Canada }

\date{\today}

\maketitle
\vspace{-1cm}

\begin{abstract}
We consider two qubits interacting with local and collective thermal reservoirs. Each spin-reservoir interaction consists of an energy exchange and an energy conserving channel. We prove a resonance representation of the reduced dynamics of the spins, valid for all times $t\geq 0$, with errors (small interaction) estimated rigorously, uniformly in time. Subspaces associated to non-interacting energy differences evolve independently, partitioning the reduced density matrix into dynamically decoupled clusters of jointly evolving matrix elements. Within each subspace the dynamics is markovian with a generator determined entirely by the resonance data of the full Hamiltonian. Based on the resonance representation we examine the evolution of entanglement (concurrence). We show that, whenever thermalization takes place, entanglement of any initial state dies out in a finite time and will not return. For a concrete class of initially entangled spin states we find explicit bounds on entanglement survival and death times in terms of the initial state and the resonance data. 
\end{abstract}

\thispagestyle{empty}
\setcounter{page}{1}
\setcounter{section}{1}


\setcounter{section}{0}


\section{Introduction and outline of main results}
\label{sec:introduction}

We consider two qubits $\s_1$, $\s_2$ (two spins $\frac12$) in a thermal environment. If the spatial separation of the qubits is small
compared to the correlation length of the environment then the qubits feel {\it the same} interaction with the reservoir, called a {\it collective} interaction. For separated qubits the interaction is
described by {\it independent} reservoirs coupled to each qubit, called a {\it local}
interaction. In experiments both kinds of interaction occur
at the same time and so we consider three independent heat
reservoirs, $\r_1, \r_2$ (local) and $\r$ (collective). 

Each of the interactions $\s_1+\r_1$, $\s_2+\r_2$ and 
$(\s_1,\s_2)+\r$ has two channels given by {\it energy
conserving} and {\it energy exchange} couplings. The former are
also called ``non-demolition'' interactions, leaving the energy of $\s_1$ and $\s_2$
invariant. Under this evolution the populations are constant (diagonal elements of reduced density matrix of
$\s_1+\s_2$ in the energy basis). Nevertheless, correlations decay in time (off-diagonal density matrix
elements). This process is called ``phase
decoherence''. Models with constant populations have a multitude of invariant states (at least all diagonal density matrices). They cannot describe asymptotic processes such as the approach to a final
(equilibrium) state. Those are induced by energy exchange
interactions. This is why both interaction channels should be considered simultaneously. 

\medskip
Our {\bf main results} are Theorems \ref{thm01} and \ref{thmrates} on the reduced dynamics, and Theorem \ref{thm3} on entanglement evolution. In the former we prove a ``resonance approximaton'' of the reduced dynamics with a remainder small in the interaction, valid {\it uniformly in time $t\geq 0$}. Theorem \ref{thm01} shows that the dynamics is expressed by just a few parameters (the resonance data). In Theorem \ref{thmrates} (and Section \ref{sectlso1}) we calculate explicitly the resonance data for the model of two qubits interacting with local and collective reservoirs. We use these results to establish, in Theorem \ref{thm3},  rigorous bounds on the times of survival and death of entanglement for a class of initial qubit states.

\medskip
{\bf Reduced dynamics.\ }  The dynamics of the qubits is obtained by tracing out the reservoirs degrees of freedom, and is given by a time-dependent reduced density matrix $\rho_t$. Assuming that the qubits are initially not entangled with the reservoirs (but may be entangled among themselves), the initial qubit state $\rho_0$ evolves according to a linear mapping of density matrices \cite{BreuerPetruccione}
$$
\rho_0\mapsto \rho_t=V_\varkappa(t)\rho_0.
$$ 
In this introduction we consider the coupling strenghts of all interactions to be proportional to some overall coupling constant $\varkappa$.

The dynamical map $V_\varkappa(t)$ is not a (semi-) group in $t$, but it is common in physics to make a so-called markovian master equation approximation (Born-, Markov- and rotating wave approximations), which restores the group property: $V_\varkappa(t)\approx \e^{t{\cal L}_\varkappa}$. Here, ${\cal L}_\varkappa ={\cal L}_0+\varkappa^2 K^\sharp$ is a Lindblad generator which includes the uncoupled dynamics and a second-order correction term (traditionally denoted by $K^\sharp$ (Davies)). The validity of this approximation is based on physical considerations and is argued to work well if the bath correlation times are much smaller than the system relaxation time \cite{BreuerPetruccione}. However, to our knowledge, no rigorous control of the error has been achieved in this method. In the extensive physics literature one finds different forms of  ${\cal L}_\varkappa$ (derived heuristically), only one of them -- the so-called Davies generator \cite{Da1,Da2} -- giving a positivity-preserving dynamics \cite{DS} (guaranteeing that $\e^{t{\cal L}_\varkappa}\rho_0$ is a density matrix for all $t\geq 0$). The mathematical procedure leading to this correct generator is the so-called weak coupling-, or Van Hove limit. It is stated as follows \cite{Da1,Da2,JP0,DJ}: for any $a>0$,
$$
\lim_{\varkappa\rightarrow 0} \sup_{\varkappa^2 t\in(0,a)} \| V_\varkappa(t) -\e^{t{\cal L}_\varkappa}\| =0.
$$
(It is assumed that $\rho$ acts on a finite dimensional Hilbert space.) 
The disadvantage of this approach is that one can only consider time scales up to $O(\varkappa^{-2})$. In \cite{FLP1,FLP2} the weak coupling analysis is extended to time scales up to $O(\varkappa^{-4})$ by replacing the semigroup $e^{t{\cal L}}$ by a more complicated mapping taking into account non-Markovian effects. Still, in order to examine large times, one has to diminish simultaneously the value of the coupling constant. We prove in Theorem \ref{thm01} a resonance approximation, improving the weak-coupling result to 
$$
\sup_{t\geq 0} \| V_\varkappa(t) -\e^{tM_\varkappa}\| \leq C\varkappa^2,
$$
where $M_\varkappa={\cal L}_0+\varkappa^2K^\sharp+\ldots$ is analytic in $\varkappa$. The gain is control uniform in $t\geq 0$, it is achieved by replacing ${\cal L}_\varkappa$ by the more complicated generator $M_\varkappa$ which contains all orders of $\varkappa$. $M_\varkappa$ commutes with the free generator ${\cal L}_0$, and the second-order correction $K^\sharp$ is the Davies generator. The latter is linked to quantum resonance theory by the relation (Proposition \ref{propos})
$$
K^\sharp = \bigoplus_{e\in {\rm spec}({\cal L}_0)} (\i\Lambda_e)^*.
$$
Here, $(\Lambda_e)^*$ is the adjoint operator of the {\it level shift operator} $\Lambda_e$ (acting on $l^1(\h_\s)=\h_\s\otimes\h_\s$, see \fer{lso}), whose spectrum yields the complex energy corrections (second order) of the unperturbed Bohr energy $e\in\mathbb R$ (energy difference of the system Hamiltonian). The eigenvalues of $M_\varkappa$ are the resonances (complex eigenvalues) of a non-selfadjoint Liouville operator (see \fer{z1}).

In the master equation setting, the {\it rotating wave} (or {\it secular}) approximation consists in neglecting, in the evolution equation, quickly oscillating terms proportional to $\e^{\i (e-e')t}$ if $e\neq e'$, where $e, e'$ are Bohr energies \cite{BreuerPetruccione}. This leads to an approximate dynamics in which spectral subspaces associated to different Bohr energies evolve independently. The fact that $M_\varkappa$ leaves eigenspaces of ${\cal L}_0$ invariant gives thus a proof of the validity of the rotating wave approximation. 
This decoupling of evolution of spectral subspaces may be very useful in the analysis of open systems with many degrees of freedom, as the density matrix is partitioned into independent clusters of jointly evolving matrix elements. Some clusters may decay quickly and one is soon left with only a `sparse' density matrix. Also, for certain applications (quantum protocols) only a few clusters may be important, and one can focus on their analysis directly by studying the corresponding level shift operators.

The present resonance approach builds on \cite{MSB1,MSB2,MSB3}, giving the dynamics of reduced density matrix elements directly in terms of spectral data. An improvement of the weak coupling limit to all times for a single spin coupled to a Bosonic reservoir has also given in \cite{JP0}.

\medskip
{\bf Evolution of entanglement.\ } Entanglement is a central notion of modern quantum theory and particularly quantum information and computation. It is a measure for quantum correlations between subsystems. In the simplest setting, entanglement measures how far away from being a product state a given state of a bipartite system is. (A product state also being called disentangled.) There are various notions of entanglement \cite{Horodecki}. We concentrate on entanglement of formation \cite{Bennet}, defined for the density matrix $\rho$ of a bipartite system $A+B$ to be
$$
{\cal E}(\rho) =\inf_{\{p_j,\psi_j\}}\sum_j p_j S\big( {\rm Tr}_B |\psi_j\rangle\langle\psi_j| \big),
$$
where $S(x)=-{\rm Tr}(x\ln x)\geq 0$ is the von Neumann entropy (when dealing with systems of spins $\frac12$ it is natural to take the binary logarithm). Here, ${\rm Tr}_B$ is the partial trace over system $B$, and the infimum is taken over all probabilities $0\leq p_j\leq 1$, $\sum_jp_j=1$, and all vector states $\psi_j$ of $A+B$, $\|\psi_j\|=1$, such that $\rho = \sum_j p_j |\psi_j\rangle\langle\psi_j|$. 

The expression for entanglement of formation involves a hard problem of minimization over all possible realizations of a given state $\rho$, and its calculation is a (too) difficult enterprise. However, Wootters \cite{wootters} has shown that if both $A$ and $B$ are spins $\frac12$ (each having state space $\cx^2$), then ${\cal E}(\rho)$ is related strictly monotonically to the {\it concurrence} $C(\rho)$ defined by an expression obtainable directly from the density matrix $\rho$, see \fer{60}-\fer{63.1}. The concurrence of two spins satisfies $0\leq C(\rho)\leq 1$. It vanishes if and only if $\rho$ can be written as a mixture (convex combination) of pure product states.

The link between entanglement and concurrence established by Wootters has provoked a wealth of investigations on entanglement of two spins in interaction with noises (classical or quantum). In particular, the phenomena of {\it entanglement decay}, {\it sudden entanglement death}, {\it death and revival} as well as {\it creation} of entanglement due to a collective reservoir have been discovered. There is a large body of work also concerning entanglement of continuous variable systems (bosonic modes), using different entanglement measures, such as ``negativities''. We refer to the review \cite{AI} and the extensive bibilography therein. In the present work, we focus on two qubits and concurrence as a measure of entanglement.

Other than numerically, entanglement has been studied in the weak coupling markovian master equation regime (Lindblad dynamics) or for explicitly solvable models. Within the context of the markovian description, once the Lindblad dynamics is assumed, a mathematically rigorous analysis is sometimes be possible. Our goal here is to start with the full microscopic (Hamiltonian) description and extract the dynamics of entanglement rigorously, without assuming a Lindblad evolution as a starting point. We next point out some references on entanglement evolution; the literature on the topic is huge and many more references are found in the papers cited below.

In the markovian approximation, Yu and Eberly \cite{YE04} find an upper bound on decay of concurrence of two qubits: $C(\rho_t)\leq \e^{-\gamma t}C(\rho_0)$ for some $\gamma>0$. (Each spin is coupled locally to a zero-temperature cavity reservoir through an energy exchange interaction.) They show that some initial states $\rho_0$ satisfy $C(\rho_t)=0$ for all times exceeding an {\it entanglement death time}, but also that there are initial states for which $C(\rho_t)>0$ for all times. Furthermore, in some models with purely energy-conserving interaction, they find {\it entanglement-free subspaces} \cite{YE03}. (Explicitly solvable model with classical (commutative, stochastic) local and collective noises.)

Bellomo et al. \cite{Bellomo} consider two spins locally interacting with zero temperature cavities in a {\it non-markovian regime} (explictly solvable energy-exchange interaction). They show that entanglement of certain initial states undergoes {\it sudden death and revival}: initial entanglement decays to zero, may stay zero for a while, but then reappears (with some loss), and so on. The interpretation is that initial entanglement is shifted from the spins to the reservoirs (intially unentangled), then shifted back to the spins with some loss. 

Braun \cite{Braun} considers two spins coupled collectively to a single thermal reservoir (energy-conserving interaction, explicitly solvable). He shows that certain initially unentangled states of the spins will acquire some entanglement for a while, due to the interaction with the common reservoir. In a similar spirit, using the Peres-Horodecki (partial transposition) criterion for entanglement detection \cite{Peres,Horodecki}, Benatti et al. show that entanglement can be created by a collective environment for certain initial conditions in markovian, not explicitly solvable models \cite{BFP,BLN}.

Our present contribution are rigorous bounds on entanglement survival and death times. We show in Theorem \ref{finitet} that the entanglement of any initial state will decay in a finite time if the entire system has the property of return to equilibrium. This property holds generically, and so the question is: how long can initial concurrence prevail, and when will it certainly have died out without returning? In Theorem \ref{thm3} we give bounds on entanglement survival and death times, linking them to the resonance data for the model with all interactions present. To our knowledge, this is the first time rigorously established bounds are given for not explicitly solvable models. 

We have announced Theorems \ref{thm01}, \ref{thmrates} and \ref{thm3} in \cite{MBBG} without proofs. The main result of that paper is a numerical analysis showing that the resonance approximation captures the phenomena of sudden death and revival, and of creation of entanglement.


\section{Main results}

\subsection{Model}

The space of pure states of $\s=\s_1+\s_2$ is given by
\begin{equation}
\h_\s = \cx^2\otimes\cx^2,
\label{1}
\end{equation}
its Hamiltonian is
\begin{equation}
H_\s = B_1S^z_1+B_2S_2^z.
\label{2}
\end{equation}
Here $B_1,B_2>0$ are the values of an (effective) magnetic field
at the positions of the two spins,
$$
S^z=
\left[
\begin{array}{cc}
1 & 0\\
0 & -1
\end{array}
\right]
$$
is a Pauli matrix and $S_1^z=S^z\otimes\bbbone_{\cx^2}$,
$S_2^z=\bbbone_{\cx^2}\otimes S^z$.

Each thermal reservoir is described by a spatially infintely
extended free bose gas in equilibrium at temperature
$1/\beta>0$,\footnote{Our approach and our results can easily be
applied to the situation where each of the three reservoirs has a
different temperature, but the scope of this work is to compare
local versus collective phenomena, and thus we take all reservoir
temperatures to be the same.} having positive particle density (phase without Bose-Einstein condensate). A
mathematical description of reservoirs of free particles is usually given
in terms of a state of the CCR Weyl algebra \cite{BR, AW}, but we explain our model here in terms of $a^*(k), a(k)$, the bosonic
creation and annihilation operators of a particle with momentum
$k\in \rx^3$, satisfying the canonical commutation relation $[a^*(k),a(l)] =\delta(k-l)$. The operators $a(k)$, $a^*(k)$ act on bosonic Fock space (represented here in Fourier, or momentum space) 
\begin{equation}
{\cal F} = \bigoplus_{n=0}^\infty L^2_{\rm sym}(\rx^{3n},\d^{3n}k),
\label{fockspace}
\end{equation}
where $L^2_{\rm sym}$ is the space of all functions invariant under permutation of the arguments $k_1,\ldots,k_n$ (symmetric functions).
The Hamiltonian of the reservoir is given by
\begin{equation}
H_\r = \int_{\rx^3} |k| a^*(k)a(k)\d^3k,
\label{8}
\end{equation}
where $|k|$ is the energy of a single particle with momentum $k$. The equilibrium state of a reservoir at temperature $1/\beta>0$ is the quasi-free state $\omega_\beta$ determined by
\begin{equation}
\omega_\beta(a(k)) = \omega_\beta(a^*(k))=0, \qquad \omega_\beta(a^*(k)a(l)) = \frac{\delta(k-l)}{e^{\beta|k|}-1}.
\label{13}
\end{equation}
The second relation implies that the probability density distribution of particles with momentum $k$ is given by $(\e^{\beta|k|}-1)^{-1}$, which is with {\it Planck's black-body radiation law}.

For $g\in L^2(\rx^3,\d^3k)$ we define the smoothed-out creation and annihilation operators
$$
a^*(g) = \int_{\rx^3} g(k) a^*(k)\d^3k, \qquad a(g) = \int_{\rx^3} \overline{g(k)} a(k)\d^3k,
$$
where $\overline{g(k)}$ is the complex conjugate of $g(k)$ and we define the field operator
$$
\phi(g) =\frac{1}{\sqrt 2}\left[ a^*(g) +a(g)\right].
$$

The Hilbert space of the total system is given by the tensor
product
\begin{equation}
\h = \h_\s\otimes {\cal F}_{\r_1}\otimes{\cal
F}_{\r_2}\otimes{\cal F}_{\r}, \label{9}
\end{equation}
where each ${\cal F}_{\r_1}, {\cal F}_{\r_2}, {\cal F}_{\r}$ is a
copy of $\cal F$, and $\h_\s$ is given in \fer{1}. The full
Hamiltonian takes the form
\begin{eqnarray}
H &=& H_\s +H_{\r_1}+H_{\r_2} + H_{\r} \label{3}\\
&& +\left( \lambda_1 S^x_1 +\lambda_2 S^x_2\right)\otimes\phi(g)\label{4}\\
&& +\left( \kappa_1 S^z_1 +\kappa_2 S^z_2\right)\otimes\phi(f)\label{5}\\
&& + \,\mu_1 S^x_1\otimes\phi(g_1) +\mu_2 S^x_2\otimes\phi(g_2)\label{6}\\
&& + \,\nu_1 S^z_1\otimes\phi(f_1) +\nu_2 S^z_2\otimes\phi(f_2)\label{7}.
\end{eqnarray}
The first four terms, \fer{3}, are the uncoupled Hamiltonians,
where $H_\s$ is given in \fer{2} and each of the $H_{\r_1},
H_{\r_2}, H_{\r}$ is the operator \fer{8}, acting on the
appropriate factor of the Hilbert space \fer{9}.

The {\it collective interaction} is determined by the operators
\fer{4} and \fer{5}, where $\phi(g)$ acts nontrivially only on the
factor ${\cal F}_{\r}$ in \fer{9}. The terms \fer{4} describe
collective energy exchange interactions of the spins with $\r$,
with respective coupling constants $\lambda_1, \lambda_2\in\rx$.
Energy exchanges are implemented by the spin flip operator (Pauli
matrix)
$$
S^x=
\left[
\begin{array}{cc}
0 & 1\\
1 & 0
\end{array}
\right],
$$
acting on either of the factors $\cx^2$ of $\h_\s$ (denoted by $S^x_1$ or $S^x_2$, analogously to $S^z_1$, $S^z_2$ in \fer{2}).
The operator \fer{5} commutes with $H_\s$ and describes collective energy preserving interactions, with respective coupling constants $\kappa_1,\kappa_2$ and field operator $\phi(f)$ acting nontrivially on the factor ${\cal F}_\r$ of $\h$.

The {\it local interactions} are governed by the operators \fer{6}
(local, energy exchange) and \fer{7} (local, energy conserving).
Again, $\mu_1, \mu_2, \nu_1, \nu_2\in\rx$ are coupling constants.
The field operators $\phi(g_j)$ and $\phi(f_j)$, $j=1,2$, act
nontrivially on the factor ${\cal F}_{\r_j}$ of $\h$.

\bigskip

For any state  $\omega_\s$ on $\s=\s_1+\s_2$  (positive linear functional on the algebra of bounded operators ${\cal B}({\cal H}_\s)$) and any system
observable $A\in{\cal B}(\h_\s)$, we denote by $\omega_\s^t(A)$
the reduced dynamics at time $t$, given by
\begin{equation}
\omega_\s^t(A) =\omega_\s\otimes\omega_{\beta,\r_1}\otimes
\omega_{\beta,\r_2}\otimes\omega_{\beta,\r}\left( \e^{\i tH}
(A\otimes\bbbone_{\r_1}
\otimes\bbbone_{\r_2}\otimes\bbbone_{\r})\e^{-\i tH}\right),
\label{40}
\end{equation}
where $H$ is the full Hamiltonian \fer{3}-\fer{7}. This expression is formal, and we understand that the termodynamic limit of all reservoirs is taken.\footnote{A rigorous discussion of this point, for a general
finite-dimensional system $\s$ is coupled linearly in the field
operators is given in \cite{MSB1, MSB2}.
}
The system dynamics \fer{40} determines the reduced density matrix $\rho_t$ by $\omega_\s^t(A) ={\rm Tr}_\s(\rho_t A)$.

\subsection{Results}

Our main results on decoherence and disentanglement are based on a
careful analysis of the reduced dynamics of $\s=\s_1+\s_2$, based on a refinement of the dynamical theory of quantum resonances developed in
\cite{MSB1, MSB2}. This theory uses analytic spectral
deformation methods and requires the following regularity
assumption.
\begin{itemize}
\item[(A)] Write $h(r,\Sigma)$ for a function $h\in L^2(\rx^3,\d^3k)$ in spherical coordinates $(r,\Sigma)\in\rx_+\times S^2$, and consider the function
\begin{equation}
\rx\times S^2\ni (u,\Sigma)\mapsto h_\beta(u,\Sigma):=\sqrt{\frac{u}{1-\e^{-\beta u}}}\, |u|^{1/2} \left\{
\begin{array}{ll}
h(u,\Sigma), & \mbox{if $u\geq 0$}\\
-\e^{-\i\alpha}\overline{h(-u,\Sigma)}, & \mbox{if $u<0$}
\end{array}
\right.
\label{10}
\end{equation}
where $\alpha\in\rx$ is any fixed phase, and $\beta>0$ is the inverse temperature. It is assumed that for $h=g, g_1, g_2, f, f_1, f_2$ (see \fer{4}-\fer{7}), the function $\theta\mapsto h_\beta(u+\theta,\Sigma)$ has an analytic extension in the variable $\theta$, as a map $\cx\times S^2\mapsto L^2(\rx\times S^2,\d u\d\Sigma)$, from $\theta\in\rx$ to $\theta\in\{ z\in \cx\ :\ 0< \Im z <\theta_0\}$, such that the extension is continuous as $\Im\theta\downarrow 0$. Here, $\theta_0>0$ is an arbitrary constant bounded above by $2\pi/\beta$ (singularity of the square root in \fer{10}).
\end{itemize}

A family of form factors $h$ satisfying this condition is $h(r,\Sigma) = r^p\e^{-r^m}h_1(\Sigma)$, with $p=-1/2+n$, $n=0,1,\ldots$ and $m=1,2$, and where $h_1$ is an arbitrary (integrable) function of the angle $\Sigma$. This family contains the usual physical form factors, \cite{PSE}.

\subsubsection{Dominant reduced dynamics}

Denote the eigenvalues of $H_\s$ by
\begin{equation}
E_1=B_1+B_2,\ E_2=B_1-B_2,\ E_3=-B_1+B_2,\ E_4=-B_1-B_2,
\label{46}
\end{equation}
with corresponding ordered basis $\{\varPhi_1,\ldots,\varPhi_4\}$
of $\cH_\s$. Explicitly,
\begin{equation}
\varPhi_1 = |++\rangle,\ \varPhi_2 = |+-\rangle,\  \varPhi_3 = |-+\rangle,\ \varPhi_4 = |--\rangle,\ 
\label{Sbasis}
\end{equation}
where $|\sigma_1\sigma_2\rangle = |\sigma_1\rangle\otimes|\sigma_2\rangle$ and $S^z|\pm\rangle =\pm|\pm\rangle$. Define
\begin{equation}
\varkappa:= \max\{ |\kappa_j|, |\lambda_j|, |\mu_j|, |\nu_j|\ :\
j=1,2\}. \label{12}
\end{equation}
Under the non-interacting dynamics ($\varkappa=0$), the evolution
of the reduced density matrix elements of $\s$ (expressed in the
energy basis \fer{Sbasis}) is given by
\begin{equation}
[\rho_t]_{kl} = \scalprod{\varPhi_k}{\e^{-\i tH_\s}\rho_0\, \e^{\i
tH_\s}\varPhi_l} = \e^{\i t e_{lk}} [\rho_0]_{kl}, \label{30}
\end{equation}
where $e_{lk}=E_l-E_k$. As the interactions with the reservoirs
are turned on (some of $\kappa_j,\lambda_j,\mu_j,\nu_j$ nonzero),
the dynamics \fer{30} undergoes two qualitative changes.
\begin{itemize}
\item[--] The ``Bohr frequencies''
\begin{equation}
e\in \{ E-E'\ : E,E'\in{\rm spec}(H_\s)\}
\label{17}
\end{equation}
in the exponent of \fer{30} become {\it complex resonance energies}, $e\mapsto
\varepsilon_e$ with non-negative imaginary parts. If $\Im\varepsilon_e >0$ then the
corresponding dynamical process is irreversible (decay).
 \item[--] Matrix elements do not evolve independently since the
effective energy of $\s$ is changed due to the interaction with
the reservoirs.
\end{itemize}

\noindent
 Both effects are small if the coupling is
small. In particular, $\varepsilon_e\rightarrow e$ as
$\varkappa\rightarrow 0$. In order to set up a perturbation theory
we view the energy differences \fer{17} as the eigenvalues
of the {\it Liouville operator}
\begin{equation}
L_\s = H_\s\otimes\bbbone_{\s} - \bbbone_{\s}\otimes H_\s,
\label{16'}
\end{equation}
acting on the doubled Hilbert space $\h_\s\otimes\h_\s$. Denote
the spectral projection of $L_\s$ associated to $e$ by $P_e$. We
have $\dim P_e={\rm mult}(e)$ (multiplicity of eigenvalue).
Generally, as the coupling parameters are turned on, there is a
multitude of distinct resonance energies $\varepsilon_e^{(s)}$
bifurcating out of $e$, with $s=1,\ldots,\nu(e)$ and $\nu(e)\leq
{\rm mult}(e)$.  The shift of eigenvalues $e$ under perturbation
is of order at least two in the coupling constants in our
model.\footnote{This follows from the fact that the coupling
operators \fer{4}-\fer{7} are linear in the field operators, thus
having zero average in the thermal state, which implies that the
first order corrections to the energies vanish as well.}  The lowest order corrections $\delta_e^{(s)}$ are $O(\varkappa^2)$. They are the eigenvalues of
an explicit {\em level shift operator} $\Lambda_e$ (see
\fer{lso}), acting on ${\rm Ran}P_e$: there are two 
bases $\{\eta_e^{(s)}\}$ and $\{\widetilde\eta_e^{(s)}\}$ of ${\rm
Ran}P_e$, satisfying
\begin{equation}
\Lambda_e\eta_e^{(s)} = \delta_e^{(s)}\eta_e^{(s)}, \qquad [\Lambda_e]^*\widetilde\eta_e^{(s)} = \overline{\delta_e^{(s)}}\widetilde\eta_e^{(s)}, \qquad \scalprod{\widetilde\eta_e^{(s)}}{\eta_e^{(s')}} =\delta_{s,s'}. 
\label{19}
\end{equation}
Moreover, we have 
\begin{equation}
\varepsilon_e^{(s)} = e +\delta_e^{(s)} +O(\varkappa^4)
\label{u1}
\end{equation}
and $\Im\delta_e^{(s)}\geq 0$.

If all the resonance
energies $\varepsilon_e^{(s)}$ become distinct at this lowest
order of perturbation, for generic small values of the
perturbation parameters, then we say that the ``Fermi Golden Rule
Condition'' is satisfied. By ``generic small'' values of the
parameters $\kappa_j,\lambda_j,\mu_j,\nu_j$, $j=1,2$, we mean that
they belong to a small ball in $\rx^8$ around the origin, except
possibly to a set of measure zero inside the ball. (Note that the
origin is not a generic point.) This assumption, expressed as
follows, is satisfied in many applications (as in the model
of the present paper).
\begin{itemize}
\item[(F)] For generic small values of the coupling constants, there is complete splitting of resonances under perturbation at second order, i.e., all the $\delta_e^{(s)}$ are distinct for fixed $e$ and varying $s$. 
\end{itemize}
This condition implies that $\nu(e)={\rm mult}(e)$ for all $e$. To quantify the clustering of density matrix elements into groups who evolve jointly under the full evolution, we introduce for each energy difference $e$, \fer{17}, the {\it cluster set}
\begin{equation}
{\cal C}(e)=\{ (k,l)\ :\ E_k-E_l=e\}.
\label{32}
\end{equation}

Denote by $[\rho_t]_{mn}$ the element $m,n$ of the reduced density
matrix at time $t$ in the energy basis $\{\varPhi_j\}$, i.e.
\begin{equation}
[\rho_t]_{mn} = \omega^t(|\varPhi_n\rangle\langle\varPhi_m|).
\label{33}
\end{equation}

The following result is obtained from a detailed analysis of a
representation of the reduced dynamics given in \cite{MSB1, MSB2,
MSB3}. We present this result for the specific system at hand,
but mention that the proof (Section \ref{proofthm01}) relies
entirely on generic properties of resonance vectors and energies,
and the result can be proven for general $N$-level systems
coupled to heat reservoirs.

\begin{thm}[Dominant reduced dynamics]
\label{thm01} Suppo\-se that Conditions (A) and (F) hold.  There
is a constant $\varkappa_0>0$ such that if $\varkappa
<\varkappa_0$, then we have for all $t\geq 0$
\begin{equation}
[\rho_t]_{mn} =\sum_{(k,l)\in{\cal C}(E_m-E_n)}A_t(m,n;k,l)\  [\rho_0]_{kl} +O(\varkappa^2),
\label{42}
\end{equation}
where the remainder term is uniform in $t\geq 0$, and where the amplitudes $A_t$ satisfy the {\em Chapman-Kolmogorov} equation
\begin{equation}
A_{t+r}(m,n;k,l)= \sum_{(p,q)\in{\cal C}(E_m-E_n)}A_t(m,n;p,q)A_r(p,q;k,l),
\label{chko}
\end{equation}
for $t,r\geq 0$, with initial condition
\begin{equation}
A_0(m,n;k,l) = \delta_{m=k}\delta_{n=l}
\label{chko1}
\end{equation}
(Kronecker delta). Moreover, the amplitudes are given in terms of
the resonance vectors and resonance energies by
\begin{equation}
A_t(m,n;k,l) =  \sum_{s=1}^{{\rm mult}(E_n-E_m)} \e^{\i
t\varepsilon_{E_n-E_m}^{(s)}}
\scalprod{\varPhi_l\otimes\varPhi_k}{\eta_{E_n-E_m}^{(s)}}
\scalprod{\widetilde\eta_{E_n-E_m}^{(s)}}{\varPhi_n\otimes\varPhi_m}.
 \label{35}
\end{equation}
\end{thm}

\noindent
{\bf Discussion.} {\bf (1)} The result shows that to lowest order in $\varkappa$, {\em and uniformly in time}, the reduced density matrix elements evolve in clusters. A cluster is determined by indices in a fixed ${\cal C}(e)$. Within each cluster the dynamics has the structure of a classical (commutative) Markov process. In general the `transition probabilities' $A_t(m,n;k,l)$ are complex valued. The typical stochasticity property $\sum_{(k,l)\in{\cal C}(E_n-E_m)}A_t(m,n;k,l)=1$ is not satisfied. (Generally, due to decoherence, all matrix elements corresponding to the (non-diagonal) clusters decay exponentially quickly to zero at the rate $\min_s\Im \varepsilon_{E_n-E_m}^{(s)}$.) However one sees readily that if the diagonal elements form a cluster by themselves (which is the case e.g. if the kernel of $L_\s$ has dimension $\dim{\cal H}_\s$), then this cluster satisfies the stochasticity condition: $\sum_{(k,k)\in{\cal C}(0)} A_t(m,m;k,k)=1$.

The clustering implies the dynamical decoupling of eigenspaces associated to different Bohr energies. This, together with the Chapman-Kolmogorov relation, shows that each subspace has a (Markov) generator of dynamics. Hence the existence of $M_\varkappa$ and its commutativity with ${\cal L}_0$, as mentioned in the introduction.

{\bf (2)} The resonance energies
$\varepsilon_e^{(s)}$ contain, in general, terms of all (even)
orders in $\varkappa$. It may happen that, due to degeneracies of
energy levels of $H_\s$, condition (F) is not satisfied, and that
zero is still a degenerate eigenvalue at order $\varkappa^2$
\cite{MLSO}. In this situation a result similar to Theorem
\ref{thm01} can be proven.

{\bf (3)} The existence of an equilibrium state of the coupled system implies that one of the resonances $\varepsilon_0^{(s)}$ is always zero \cite{MLSO}, we set $\varepsilon_0^{(1)}=0$. The condition $\Im\varepsilon_e^{(s)}>0$ for all $e,s$ except $e=0$, $s=1$ is equivalent to the system converging to its equilibrium state, by which we mean that $\lim_{t\rightarrow\infty}\omega^t(A)=\omega_\beta(A)$ for all observables $A$ of $\s$, where $\omega_\beta$ is the state of $\s$ obtained by reducing the {\it coupled} equilibrium state of $\s$ plus the reservoirs. $\omega_\beta$ coincides with the Gibbs state of $\s$ at temperature $1/\beta$, up to $O(\varkappa^2)$.

{\bf (4)} For generic couplings, $\s$ undergoes decoherence in the energy basis, which means that the off-diagonal density matrix elements converge to zero for large times, at lowest order in the coupling. However, the diagonal elements cannot experience the same fate, since they must sum up to one at all times (${\rm Tr}\rho_t=1$). 
\begin{itemize}
\item[--] The {\it thermalization rate} is defined by
\begin{equation}
\gamma^{\rm th} = \min_{s\geq 2}\Im\varepsilon_0^{(s)}\geq 0.
\label{66}
\end{equation}
(Recall that $\varepsilon_0^{(1)}=0$, see remark (3) above.)
The {\it cluster decoherence rate} associated to ${\cal C}(e)$, $e\neq 0$, is defined to be
\begin{equation}
\gamma^{{\rm dec}}_e =\min_s \Im\varepsilon_e^{(s)} \geq 0 \qquad (1\leq s \leq {\rm mult}(e)).
\label{36}
\end{equation}
If $\gamma$ is any of the above rates, then $\tau=1/\gamma$ is the corresponding (thermalization, decoherence) time. We use the convention $\tau=\infty \Leftrightarrow \gamma=0$.

\item[--]
We say the system has the property of {\em return to equilibrium} if any normal initial state converges to the (joint) equilibrium state as time tends to infinity.\footnote{More precisely, let $\beta$ be the inverse temperature of the reservoirs $\r_1, \r_2, \r$, and consider the reference state
$\omega_{\rm ref}= \omega_{\s_1}\otimes\omega_{\s_2}\otimes\omega_{\r_1,\beta}\otimes_{\r_2,\beta}\otimes\omega_{\r,\beta}$, where $\omega_{\s_j}$ are arbitrary states on $\s_j$, $j=1,2$, and $\omega_{\r_j,\beta}$ is the $\beta$-KMS state of reservoir $\r_j$. Let $\frak M$ be the algebra of observables of the total system, represented as a von Neumann algebra acting on the GNS representation Hilbert space ${\cal H}_{\rm ref}$ of $\omega_{\rm ref}$. Let $\alpha^t$ be the Heisenberg dynamics of the total system, $\alpha^t$ is a group of $*$automorphisms of $\frak M$. The system has the property of return to equilibrium if for any state $\omega$ represented by a density matrix on ${\cal H}_{\rm ref}$ and any $A\in{\frak M}$, we have $\lim_{t\rightarrow\infty} \omega(\alpha^t(A)) = \omega_{\beta,\varkappa}(A)$, where $\omega_{\beta,\varkappa}$ is the $\beta$-KMS state  w.r.t. the coupled dynamics of the whole system.}
\end{itemize}


\subsubsection{Finite disentanglement time}

Let $\rho$ be the density matrix of two spins $\frac12$. The {\it
concurrence} \cite{wootters} is defined by
\begin{equation}
C(\rho) = \max\{ 0, D(\rho)\},
\label{60}
\end{equation}
with
\begin{equation}
D(\rho) = \sqrt{\nu_1}-\big[\sqrt{\nu_2}+\sqrt{\nu_3}+\sqrt{\nu_4}\big],
\label{61}
\end{equation}
and where $\nu_1\geq\nu_2\geq\nu_3\geq\nu_4\geq0$ are the eigenvalues of the matrix
\begin{equation}
\xi(\rho) = \rho(S^y\otimes S^y) \bar\rho(S^y\otimes S^y).
\label{62}
\end{equation}
Here, $\bar\rho$ is obtained from $\rho$ by representing the latter in the energy basis and then taking the elementwise complex conjugate, and $S^y$ is the Pauli matrix
\begin{equation}
S^y =
\left[
\begin{array}{cc}
0 & -\i \\
\i & 0
\end{array}
\right].
\label{63.1}
\end{equation}
It is rather easy to see that all eigenvalues of $\xi(\rho)$ are
non-negative (but $\xi(\rho)$ is not hermitian). The concurrence
\fer{60} takes values in the interval $[0,1]$. If $C(\rho)=0$ then
the state $\rho$ is separable, meaning that $\rho$ can be written
as a mixture of pure product states. If $C(\rho)=1$ then $\rho$ is
called maximally entangled.

\medskip
Let $\rho_0$ be an initial state of $S$. The smallest $t_0\geq 0$ s.t. $C(\rho_t)=0$ for all $t\geq t_0$ is called the {\em disentanglement time} (also `entanglement sudden death time', \cite{YE03, YE04}). If $C(\rho_t)>0$ for all $t\geq 0$ then we set $t_0=\infty$. The disentanglement time depends on the initial state.

\medskip
As mentioned in Section \ref{sec:introduction}, it is well known that under energy-conserving interactions, initial entanglement may decay gradually to zero without being zero at any finite time, or it may even stay constant. However, systems with energy exchange (where $H_\s$ is not conserved) generically have the property of return to equilibrium \cite{JPrte,BFS,FM}. A consequence of thermalization is death of entanglement (in finite time). The mechanism is very simple: the equilibrium (Gibbs) state is the centre of a neighbourhood of disentangled states, of size $\propto [{\rm Tr}\,\e^{-\beta H_\s}]^{-1}$. By the property of return to equilibrium, the qubit state approaches its final state which is the qubit Gibbs state plus a correction of the order $\varkappa^2$ (reduction of joint equilibrium of qubits and reservoirs). Thus under the condition $\varkappa^2<\!\!< [{\rm Tr}\,\e^{-\beta H_\s}]^{-1}$ the qubit state will enter and stay in the entanglement free neighbourhood. We give some estimates in Section \ref{sectthmfinitet}.

\begin{thm}[Finite disentanglement time]
\label{finitet}
Suppose the system has the property of return to equilibrium at some temperature $T=1/\beta>0$ and for some $\varkappa\neq 0$. Let $\rho_0$ be any initial state of the qubits. Then there is a constant $c$ (independent of $\beta$ varying in compact sets and of $\rho_0$) s.t. if $\varkappa^2\leq c\, [{\rm Tr}\,\e^{-\beta H_\s}]^{-1}$, then we have $C(\rho_t)=0$ (concurrence) for all $t\geq t_0$, where $t_0<\infty$ depends on $\beta,\varkappa$ and $\rho_0$.

If in addition return to equilibrium happens exponentially quickly at the rate $1/\tau_{\rm th}$, then there is a constant $c'>0$ s.t.
$ t_0\leq \tau_{\rm th} \ln[c' {\rm Tr}\, \e^{-\beta H_\s}]$.
\end{thm}

\noindent
Theorem \ref{finitet} shows that if the coupling constants are small, for fixed temperature, then the disentanglement time is finite. However, it is shown in \cite{Paz} that if, at fixed coupling constants, the temperature is decreased sufficiently, then entanglement can
persist for all times even if the system has the property of
return to equilibrium.

\subsubsection{Decoherence and thermalization rates}
\label{concmodsect}

We consider the Hamiltonian $H_\s$, \fer{2}, with parameters $0<B_1<B_2$ s.t. $B_2/B_1\neq 2$. 
These assumptions represent a non-degeneracy
condition which is not essential for the applicability of our
method, but lightens the exposition. The eigenvalues of $H_\s$
are given by \fer{46} and the spectrum of $L_\s$ is $\{e_1,\pm
e_2,\pm e_3,\pm e_4,\pm e_5\}$, with non-negative eigenvalues
\begin{equation}
e_1=0,\  e_2=2B_1,\  e_3=2B_2,\  e_4=2(B_2-B_1),\  e_5=2(B_1+B_2),
\label{45}
\end{equation}
having multiplicities $m_1=4$, $m_2=m_3=2$, $m_4=m_5=1$,
respectively. According to \fer{45}, the grouping of jointly
evolving elements of the density matrix above and on the diagonal
is given by five clusters \footnote{Since the density matrix is hermitian, it
suffices to know the evolution of the elements on and above the
diagonal.}
\begin{eqnarray}
{\cal C}_1 &:=&{\cal C}(e_1) =\{(1,1), (2,2), (3,3), (4,4)\}\label{47}\\
{\cal C}_2 &:=&{\cal C}(e_2) = \{ (1,3), (2,4)\}\label{48}\\
{\cal C}_3 &:=&{\cal C}(e_3) = \{ (1,2), (3,4)\}\label{49}\\
{\cal C}_4 &:=&{\cal C}(-e_4) = \{ (2,3)\}\label{50}\\
{\cal C}_5 &:=&{\cal C}(e_5) = \{ (1,4)\}\label{51}
\end{eqnarray}
{}For $x>0$ and $h\in L^2(\rx^3,\d^3k)$ we define
\begin{equation}
\sigma_h(x) = 4 \pi x^2 \coth(\beta x) \int_{S^2} |h(2x,\Sigma)|^2\d\Sigma,
\label{57}
\end{equation}
and for $x=0$ we set
\begin{equation}
\sigma_h(0) = 4 \pi \lim_{x\downarrow 0} x^2 \coth(\beta x) \int_{S^2} |h(2x,\Sigma)|^2\d\Sigma.
\label{58}
\end{equation}
Furthermore, let
\begin{eqnarray}
Y_2 &=&  \big| \Im \left[ 4\kappa_1^2\kappa_2^2 r^2 -\textstyle\frac14\i (\lambda_2^2+\mu_2^2)^2 \sigma^2_g(B_2) -4\i \kappa_1\kappa_2  \ (\lambda_2^2+\mu_2^2) \ r r_2'  \right]^{1/2}\big|, \label{69}\\
Y_3 &=&  \big| \Im \left[ 4\kappa_1^2\kappa_2^2 r^2 -\textstyle\frac14\i (\lambda_1^2+\mu_1^2)^2 \sigma^2_g(B_1) -4\i \kappa_1\kappa_2  \ (\lambda_1^2+\mu_1^2) \ r r_1'  \right]^{1/2}\big|,\label{70}
\end{eqnarray}
(principal value square root with branch cut on negative real axis) where 
\begin{equation}
r={\rm P.V.}\int_{\rx^3} \frac{|f|^2}{|k|}\d^3k,\qquad r_j' = 4\pi B_j^2\int_{S^2}|g(2 B_j,\Sigma)|^2 \d\Sigma.
\label{pvr}
\end{equation}

\begin{thm}[Decoherence and thermalization rates]
\label{thmrates}
Take coupling functions in \fer{4} -\fer{7} satisfying $f_1=f_2=f$, $g_1=g_2=g$. The thermalization and decoherence rates are given by
\begin{eqnarray}
\gamma^{\rm th} &=& \min_{j=1,2}\left\{ (\lambda_j^2+\mu_j^2)\sigma_g(B_j)\right\}+O(\varkappa^4) \label{52}\\
\gamma_2^{\rm dec} &=& \textstyle\frac 12 (\lambda^2_1+\mu_1^2)\sigma_g(B_1) + \textstyle\frac 12 (\lambda^2_2+\mu_2^2)\sigma_g(B_2) \nonumber\\
&& - Y_2 +(\kappa_1^2+\nu^2_1)\sigma_f(0) +O(\varkappa^4)\label{53}\\
\gamma_3^{\rm dec} &=& \textstyle\frac 12 (\lambda^2_1+\mu_1^2)\sigma_g(B_1) + \textstyle\frac 12 (\lambda^2_2+\mu_2^2)\sigma_g(B_2) \nonumber\\
&& - Y_3 +(\kappa_2^2+\nu^2_2)\sigma_f(0) +O(\varkappa^4)\label{54}\\
\gamma_4^{\rm dec} &=& (\lambda^2_1+\mu_1^2)\sigma_g(B_1) + (\lambda^2_2+\mu_2^2)\sigma_g(B_2) \nonumber\\
&& +\left[(\kappa_1-\kappa_2)^2 +\nu_1^2+\nu_2^2\right]\sigma_f(0) +O(\varkappa^4) \label{55}\\
\gamma_5^{\rm dec} &=& (\lambda^2_1+\mu_1^2)\sigma_g(B_1) + (\lambda^2_2+\mu_2^2)\sigma_g(B_2) \nonumber\\
&& +\left[(\kappa_1+\kappa_2)^2 +\nu_1^2+\nu_2^2\right]\sigma_f(0)+O(\varkappa^4) \label{56}
\end{eqnarray}
\end{thm}

\noindent
We give a proof in Section \ref{sectproofthmrates}.

{\bf Discussion.} {\bf (1)} The thermalization rate depends on energy-exchange couplings only. This is natural since energy-conserving interactions leave the populations invariant. 

{\bf (2)} For the purely energy-exchange model ($\kappa_j=\nu_j=0$) the rates depend symmetrically on the local and collective interactions trough the combination $\lambda^2_j+\mu^2_j$. However, for the purely energy-conserving interaction ($\lambda_j=\mu_j=0$) the rates are not symmetric in the local and collective interaction. For instance, $\gamma_4^{\rm dec}$ may depend on the local interaction only ($\kappa_1=\kappa_2$). 

{\bf (3)} The effect of energy-conserving and energy-exchange couplings, and of local and collective ones, are correlated. Indeed $Y_2, Y_3$ are complicated functions of all interaction parameters, except the local energy-conserving ones ($\nu_j$).

\subsubsection{Entanglement dynamics}
Consider the family of pure intital states of $\s$ given by $\rho_0=|\psi\rangle\langle\psi|$, with
\begin{equation}
\psi = \frac{a_1}{\sqrt{|a_1|^2+|a_2|^2}}\,  |++\rangle + \frac{a_2}{\sqrt{|a_1|^2+|a_2|^2}}\,|--\rangle,
\label{m1}
\end{equation}
where $|++\rangle$ etc are defined after \fer{Sbasis}, and where $a_1, a_2\in\cx$ are arbitrary. The concurrence is
\begin{equation}
C(\rho_0) =  2\frac{|\Re \,a_1\overline a_2|}{|a_1|^2+|a_2|^2}.
\label{m2}
\end{equation}
This class of initial states covers the whole range from
unentangled states (e.g. $a_1=0$) to maximally entangled states
(e.g. $a_1=a_2\in\rx$). According to Theorem \ref{thm01}, the
density matrix of $\s$ at time $t\geq 0$ is given by
\begin{equation}
\rho_t =
\left[
\begin{array}{cccc}
p_1 & 0 & 0 & \alpha\\
0 & p_2 & 0 & 0 \\
0 & 0 & p_3 & 0 \\
\overline\alpha & 0 & 0 & p_4
\end{array}
\right] +O(\varkappa^2),
\label{m3}
\end{equation}
with remainder uniform in $t$, and where $p_j=p_j(t)$ and
$\alpha=\alpha(t)$ are given by the main term on the r.h.s. of
\fer{42}. The initial conditions are
\begin{equation}
p_1(0) = \frac{|a_1|^2}{|a_1|^2+|a_2|^2},\ \ p_2(0)=p_3(0)=0,\ \ p_4(0) = \frac{|a_2|^2}{|a_1|^2+|a_2|^2},
\label{m4}
\end{equation}
and
\begin{equation}
\alpha(0) = \frac{\overline a_1 a_2}{|a_1|^2+|a_2|^2}.
\label{m5}
\end{equation}
We define

\begin{equation}
p:= p_1(0)\in [0,1]
 \label{p}
\end{equation}
and note that
\begin{equation}
p_4(0) =1-p \qquad \mbox{and}\qquad |\alpha(0)| = \sqrt{p(1-p)}.
 \label{star}
\end{equation}
In terms of $p$, the initial concurrence, \fer{m2}, is
$C(\rho_0)=2\sqrt{p(1-p)}$. There are four resonances associated
to the eigenvalue $e=0$. One is located at the origin, two have
leading (in $\varkappa$) imaginary parts given by (see \fer{ss1})
\begin{equation}
\delta_2 := (\lambda^2_1+\mu^2_1)\sigma_g(B_1), \qquad \delta_3 := (\lambda^2_2+\mu^2_2)\sigma_g(B_2),
\label{delta2}
\end{equation}
and the fourth one has leading imaginary part $\delta_2+\delta_3$. The leading term of the imaginary part of $\varepsilon_{2(B_1+B_2)}$ is (see \fer{56})
\begin{equation}
\delta_5 := \delta_2+\delta_3 +\left[(\kappa_1+\kappa_2)^2 +\nu_1^2+\nu_2^2\right]\sigma_f(0).
\label{deltas}
\end{equation}
We also define
\begin{equation}
\delta_+ := \max\{\delta_2,\delta_3\},\qquad \delta_-:=\min\{\delta_2,\delta_3\}.
\label{deltapm}
\end{equation}

\begin{thm}[Disentanglement time bounds.]
\label{thm3} Take  $p\neq 0,1$ and suppose that $\delta_2, \delta_3>0$. There is a constant $\varkappa_0>0$ (independent of\ $p$) such that for $0\neq\varkappa<\varkappa_0\sqrt{p(1-p)}$ we have the following.
\medskip

{\bf A.\ }  {\em (Entanglement death.)}   There is a constant $C_A>0$ (independent of\
$p,\varkappa$) s.t. $C(\rho_t)=0$ for all $t\geq t_A$, where
\begin{equation}
t_A:= \max\left\{ \frac{1}{\delta_5}\ln\left[C_A
\frac{\sqrt{p(1-p)}}{\varkappa^2}\right],
\frac{1}{\delta_2+\delta_3} \ln\left[C_A
\frac{p(1-p)}{\varkappa^2}\right],
\frac{C_A}{\delta_2+\delta_3}\right\}. \label{m17}
\end{equation}

{\bf B.\ } {\em (Entanglement survival.)} There is a constant $C_B>0$ (independent of $p$, $\varkappa$) s.t. $C(\rho_t)>0$ for all $t< t_B$, where
\begin{equation}
 t_B:=\min\left\{ \frac{1}{\delta_2+\delta_3}\ln[1+ C_B p(1-p)], \frac{1}{\delta_+}\ln\left[1+C_B\varkappa^2 \right], \frac{C_B}{\delta_5-\delta_-/2}
 \right\}. \label{m18}
\end{equation}
\end{thm}

\medskip
\noindent
{\bf Discussion.} {\bf (1)} The disentanglement time is finite since $\delta_2,\delta_3>0$ (which implies thermalization). If the system does not thermalize, then entanglement for certain initial conditions may stay nonzero for all times. 

{\bf (2)} The rates $\delta$ are of order $\varkappa^2$. Both  $t_A$ and $t_B$ increase with decreasing coupling strength. This is consistent with the expectation that disentanglent happens at a slower pace for small couplings (no disentanglement for $\varkappa=0$).

{\bf (3)} As functions of $p$, both $t_A$ and $t_B$ are maximal at $p=1/2$ and minimal at $p\in\{0,1\}$. This is consistent with the expectation that a maximally entangled state ($p=1/2$) keeps its entanglement for longest. Even if the initial state is disentangled ($p=0, 1$) we expect to see creation of entanglement due to the collective coupling (up to times at most $C_A/(\delta_2+\delta_3)$, see \fer{m17} with $p=0,1$). We show {\it numerically} in \cite{MBBG} that the resonance dynamics {\it does} reveal creation, as well as death and revival of entanglement.

\section{Proof of Theorem \ref{thm01}}
\label{proofthm01}

In \cite{MSB1,MSB2,MSB3} we have proven a representation of the reduced dynamics of a general
$N$-level system coupled to a heat reservoir. We outline here how to generalize it to the present situation of two spins coupled to three reservoirs. Of course, the generalization to a general $N$-level system in contact with any number of reservoirs is immediate.

Resonance energies $\varepsilon_e^{(s)}$ bifurcate out
($\varkappa\neq 0$) of the real energies of the system Liouvillian
$L_\s$ \fer{16'}, into the upper complex plane, c.f. \fer{u1}. They are the eigenvalues of
a closed (unbounded, non-normal) operator
\begin{equation}
K_{\vec\varkappa,\theta} = L_0 +\theta N +\sum_{j=1}^8\varkappa_j
W_j
 \label{z1}
\end{equation}
acting on the GNS Hilbert space
\begin{equation}
\h_{\rm GNS} = \h\otimes\h,
 \label{z6}
\end{equation}
see \fer{9}.
 Here, $L_0=L_\s+L_{\r_1}+L_{\r_2}+L_{\r_3}$ with
$L_\s=H_\s\otimes\bbbone -\bbbone\otimes H_\s$ and the $L_{\r_j}$
are defined similarly. We write $\vec\varkappa$ for the collection of
all the coupling constants in \fer{4}-\fer{7}. $\theta$ is a complex spectral
translation parameter (\cite{MSB1,MSB3,JPrte,BFS}) and $N=N_1+N_2+N_3$ is the
sum of the total number operators of each reservoir, with $N_j =
N\otimes\bbbone +\bbbone\otimes N$ acting on ${\cal
F}_{\r_j}\otimes {\cal F}_{\r_j}$, $N$ being the number operator
on Fock space $\cal F$. To ease notation, we have labelled in \fer{z1}
the interaction constants by $\varkappa_j$, and the $W_j$ are
interaction operators, obtained from \fer{4}-\fer{7} by passing to the GNS
space and adding suitable operators in the commutant -- making \fer{z1}
an operator of the `$C$-Liouvillian type'. The defining property of the $W_j$ is that 
$$
K_{\vec\varkappa,\theta}\ \Omega_\s\otimes\Omega_{\r_1,\r_2,\r}=0,
$$
where  (see also \fer{1} and \fer{Sbasis})
\begin{equation}
\Omega_\s = \frac{1}{2}\sum_{k=1}^4 \varPhi_k\otimes\varPhi_k,
\label{Strace}
\end{equation}
is the trace state of $S_1+S_2$, and where $\Omega_{\r_1, \r_2, \r}$ is the product state of (six) vacuum states $\Omega$ in each of the three (doubled) factors of the Fock spaces. We refer to \cite{MSB1}, Appendices A and B for the construction of the explicit expressions of the $W_j$. The
following result is the analogue of Theorem 4.1 of \cite{MSB1}.

\begin{thm}[Uncovering of resonances]
\label{thmz1}
    Suppose that condition (A)
 is satisfied. Fix any $\theta_1$ with $0<\theta_1<\theta_0$ (see condition (A)).
There is a constant $\varkappa_0$ (depending on $\theta_1$) s.t if
$\varkappa:=|\vec\varkappa|<\varkappa_0$ and $\theta_1<\Im\theta <
\theta_0$, then the operator $K_{\vec\varkappa,\theta}$ has only
isolated eigenvalues in the region
$\{z\in\cx : \Im z < \theta_1/2\}$. These eigenvalues, denoted $\varepsilon_e^{(s)}$, do not
depend on $\theta$ and have the expansion
 \begin{equation}
\varepsilon_e^{(s)} = e +\delta_e^{(s)} +O(\varkappa^4),
\label{z2}
\end{equation}
where $e\in {\rm spec}(L_\s)$, $1\leq s\leq {\rm mult}(e)$ and $\delta_e^{(s)}\in\cx$ satisfies 
$\Im\delta_e^{(s)}\geq 0$. Furthermore, the $\delta_e^{(s)}$ are
the eigenvalues of the {\em level shift operator}
\begin{equation}
\Lambda_e := -P_e W\bar P_e (L_0-e+\i 0)^{-1} \bar P_eW P_e
\upharpoonright_{{\rm Ran} P_e},
 \label{lso}
\end{equation}
where $P_e$ is the spectral projection of $L_0$ associated to the
eigenvalue $e$, $\bar P_e=\bbbone -P_e$, and where we have set for short 
$W=\sum_{j=1}^8\varkappa_j W_j$.
\end{thm}

Let $\Gamma$ be a simple closed contour containing all the
eigenvalues $\varepsilon_e^{(s)}$ of $K_{\vec\varkappa,\theta}$,
but not containing any of its continuous spectrum,  and let
\begin{equation}
Q = \frac{-1}{2\pi\i}\int_{\Gamma}
(K_{\vec\varkappa,\theta}-z)^{-1} \d z
 \label{z4}
\end{equation}
be the associated Riesz projection. It has dimension sixteen in our model of Section \ref{concmodsect}, which is the number of eigenvalues inside $\Gamma$, counting multiplicity.
The operator $K_{\vec\varkappa,\theta}$ is reduced by $Q$ and has
a finite-dimensional block $QK_{\vec\varkappa,\theta}Q$. We define
\begin{equation}
\widetilde V(t) = {\rm Tr}_{\r_1+\r_2+\r}\Big[P_\r\  \e^{\i
tQK_{\vec\varkappa,\theta}Q}\Big]\qquad \mbox{with}\qquad P_\r=|\Omega_{\r_1, \r_2,
\r}\rangle \langle \Omega_{\r_1, \r_2, \r}|,
 \label{z5}
 \end{equation}
where the trace is taken over the doubled Fock spaces of all the
reservoirs. For each $t\geq 0$,
$\widetilde V(t)$ is an operator acting on the GNS space $\cH_\s 
\otimes\cH_\s$.
The trace state $\Omega_\s$, \fer{Strace}, is cyclic and separating for the von Neumann algebra 
$\mm_\s={\cal B}(\h_\s)\otimes\bbbone\subset {\cal
B}(\h_\s\otimes\h_\s)$, and so the relation
 \begin{equation}
(V(t)A)\otimes\bbbone \ \Omega_\s = \widetilde V(t)
(A\otimes\bbbone) \Omega_\s\qquad \forall A\in{\cal B}(\h_{\s})
 \label{z7}
\end{equation}
defines uniquely a linear operator $V(t)$ acting on ${\cal B}(\h_\s)$ (Heisenberg evolution).

Theorem 3.1 of \cite{MSB1} can be formulated in the following way.

\begin{thm}[\cite{MSB1}]
\label{thmz2} Assume the conditions of Theorem \ref{thmz1}. Then there
are constants $\varkappa_0$, $C$ s.t. if
$\varkappa\leq\varkappa_0$ then
 \begin{equation}
\left| \omega_\s(\alpha^t(A)) - \omega_\s(V(t)A)\right| \leq C
\varkappa^2 \e^{-\gamma t},
  \label{z3}
 \end{equation}
for all $t\geq 0$ and all $A\in{\cal B}(\cH_\s)$. Here,
$\gamma=\Im\theta -O(\varkappa)$ satisfies
$\gamma>2\Im\varepsilon_e^{(s)}$ for all $e,s$.
\end{thm}

\subsection{Proof of Theorem \ref{thm01}}

The dynamical map $\widetilde V(t)$, \fer{z5}, does not have the
group property in general [which would mean $\widetilde V(t)\widetilde V(s)=\widetilde V(t+s)$] and hence nor does $V(t)$. However, due to assumption
(F) saying that all resonance energies are simple, we have the diagonal representation 
\begin{equation}
\e^{\i tQK_{\vec\varkappa,\theta}Q} = \sum_{e}\sum_{s=1}^{{\rm
mult}(e)} \e^{\i t \varepsilon_e^{(s)}}
|\chi_e^{(s)}\rangle\langle\widetilde\chi_e^{(s)}|,
 \label{z11}
\end{equation}
where the double sum is over all resonances (see Theorem
\ref{thmz1}), and where $K_{\vec \varkappa,\theta}
\chi_e^{(s)}=\varepsilon_e^{(s)}\chi_e^{(s)}$ and $[K_{\vec
\varkappa,\theta}]^*
\widetilde \chi_e^{(s)}=\overline{\varepsilon_e^{(s)}}\widetilde \chi_e^{(s)}$ (adjoint
operator), with normalization
\begin{equation}
\scalprod{\chi_e^{(s)}}{\widetilde\chi_{e'}^{(s')}} =\delta_{e,e'}
\delta_{s,s'}
 \label{z12}
\end{equation}
(Kronecker deltas).  An expansion in $\varkappa$ yields
\begin{equation}
|\chi_e^{(s)}\rangle\langle\widetilde\chi_e^{(s)}|=
|\eta_e^{(s)}\rangle\langle\widetilde\eta_e^{(s)}|\otimes P_\r
+\widetilde O(\varkappa), 
\label{z9}
\end{equation}
where the $\eta_e^{(s)}$ and $\widetilde\eta_e^{(s)}$ satisfy
\fer{19}, $P_\r$ is defined in \fer{z5} and where the remainder term satisfies $P_\r
\widetilde O(\varkappa)=O(\varkappa^2)$. It follows from \fer{z5}, \fer{z11}
and \fer{z9} that
\begin{equation}
\widetilde V(t) = \sum_{e}\sum_{s=1}^{{\rm mult}(e)} \e^{\i t
\varepsilon_e^{(s)}} \Big[
|\eta_e^{(s)}\rangle\langle\widetilde\eta_e^{(s)}|
+O(\varkappa^2)\Big].
 \label{z12.1}
\end{equation}
Next, we have (recall that the $\varPhi$ are given in \fer{Sbasis} and $\Omega_\s$ in \fer{Strace})
\begin{eqnarray}
\lefteqn{ |\eta_e^{(s)}\rangle\langle\widetilde\eta_e^{(s)}|
\ (|\varPhi_n\rangle\langle\varPhi_m|\otimes\bbbone)\Omega_\s
}\nonumber \\
&=&  {\textstyle \frac12} |\eta_e^{(s)}\rangle\scalprod{\widetilde\eta_e^{(s)}}{
\varPhi_n\otimes\varPhi_m}\nonumber\\
&=& {\textstyle \frac12} \sum_{k,l}
\varPhi_l\otimes\varPhi_k\scalprod{\varPhi_l \otimes\varPhi_k
}{\eta_e^{(s)}} \scalprod{\widetilde\eta_e^{(s)}}{
\varPhi_n\otimes\varPhi_m} \nonumber\\
&=&\sum_{k,l} (|\varPhi_l\rangle\langle\varPhi_k|\otimes
\bbbone)\Omega_\s \ \scalprod{\varPhi_l \otimes\varPhi_k
}{\eta_e^{(s)}} \scalprod{\widetilde\eta_e^{(s)}}{
\varPhi_n\otimes\varPhi_m}. 
\label{*}
\end{eqnarray}
Combining \fer{z12.1} and \fer{*} with \fer{z7}, we obtain
\begin{eqnarray}
\lefteqn{ V(t) |\varPhi_n\rangle\langle\varPhi_m| }\nonumber\\
&=& \sum_e\sum_{s=1}^{{\rm mult}(e)} \e^{\i t\varepsilon_e^{(s)}}
\left[ \sum_{k,l}\scalprod{\varPhi_l \otimes\varPhi_k
}{\eta_e^{(s)}} \scalprod{\widetilde\eta_e^{(s)}}{
\varPhi_n\otimes\varPhi_m} |\varPhi_l\rangle\langle\varPhi_k| +
O(\varkappa^2)\right].\qquad
 \label{z15}
\end{eqnarray}
Note that $\eta_e^{(s)}, \widetilde\eta_e^{(s)}\in{\rm Ran}
P_e$ (spectral projection of $L_\s$ associated to $e$) and therefore the main term of the sum vanishes unless $e=E_l-E_k=E_n-E_m$. Thus
\begin{eqnarray}
\lefteqn{ \omega_\s\left(V(t)|\varPhi_n\rangle\langle
\varPhi_m|\right)} \nonumber\\
&&= \sum_{s=1}^{{\rm mult}(E_n-E_m)} \e^{\i
t\varepsilon_{E_n-E_m}^{(s)}}\!\!\!\!\sum_{(k,l)\in{\cal
C}(E_m-E_n)} \scalprod{\varPhi_l \otimes\varPhi_k }{\eta_{E_n-E_m}^{(s)}}
\scalprod{\widetilde\eta_{E_n-E_m}^{(s)}}{ \varPhi_n\otimes\varPhi_m}\times\nonumber\\
&&\ \ \ \ \times\omega_\s(|\varPhi_l\rangle\langle\varPhi_k|) \ +O(\varkappa^2)\nonumber\\
&&=\sum_{(k,l)\in{\cal C}(E_m-E_n)} A_t(m,n;k,l)\  [\rho_0]_{kl}
+O(\varkappa^2),
 \label{z19}
\end{eqnarray}
where $A_t$ is given in \fer{35}. Combining \fer{z19} with Theorem
\ref{thmz2} yields \fer{42}.

Finally we check the Chapman-Kolmogorov equation \fer{chko}. Write
$E_{ij}$ for $E_i-E_j$ and $\varPhi_{ij}$ for
$\varPhi_i\otimes\varPhi_j$. The r.h.s. of \fer{chko} equals
\begin{eqnarray}
\lefteqn{ \sum_{(p,q)\in{\cal C}(E_{mn})} \sum_{s=1}^{{\rm
mult}(E_{nm})}\sum_{ s'=1}^{{\rm mult}(E_{qp})} \e^{\i
t\varepsilon_{E_{nm}}^{(s)} +\i
r\varepsilon_{E_{qp}}^{(s')}}}\nonumber\\
&&\qquad\times \scalprod{\varPhi_{qp}}{\eta_{E_{nm}}^{(s)}}
\scalprod{\widetilde\eta_{E_{nm}}^{(s)}}{\varPhi_{nm}}
\scalprod{\varPhi_{lk}}{\eta_{E_{qp}}^{(s')}}
\scalprod{\widetilde\eta_{E_{qp}}^{(s')}}{\varPhi_{qp}}.
 \label{z20}
\end{eqnarray}
Since $(p,q)\in{\cal C}(E_{mn})$ we have $E_{qp}=E_{nm}\equiv e$,
and we can perform the sum over $p,q$ in \fer{z20},
\begin{equation}
\sum_{(p,q)\in{\cal C}(E_{mn})}
\scalprod{\widetilde\eta_e^{(s')}}{\varPhi_{qp}}
\scalprod{\varPhi_{qp}}{\eta_e^{(s)}} = \scalprod{\widetilde
\eta_e^{(s')}}{\eta_e^{(s)}} =\delta_{s,s'}, \label{z21}
\end{equation}
where we use $\sum_{(p,q)\in{\cal C}(E_{nm})}|\varPhi_{qp}\rangle\langle \varPhi_{qp}| = P_e$ in the first step, and \fer{19} in the second one. Therefore, \fer{z20} becomes
\begin{eqnarray}
\sum_{s=1}^{{\rm mult}(E_{nm})} \e^{\i
(t+r)\varepsilon_{E_{nm}}^{(s)} }
\scalprod{\varPhi_{lk}}{\eta_{E_{nm}}^{(s)}}
\scalprod{\widetilde\eta_{E_{nm}}^{(s)}}{\varPhi_{nm}},
 \label{z22}
\end{eqnarray}
which is $A_{t+r}(m,n;k,l)$. Thus \fer{chko} is proven. It is also
easy to establish \fer{chko1}. This completes the proof of Theorem
\ref{thm01}. \hfill $\blacksquare$

\section{Level shift operators, proof of Theorem \ref{thmrates}}
\label{sectproofthmrates}

\subsection{Level shift operators}
\label{sectlso1}

The general form of the level shift operator (defined in \fer{lso}) with interaction linear in creation and annihilation operators has been given in \cite{MSB1}, Proposition 5.1.\footnote{In the present work we use the trace state for the two spins as reference state \fer{Strace}, while in \cite{MSB1} the Gibbs state was used -- the corresponding modification in the level shift operator is obtained simply by setting $\beta_{\s}=0$. Furthermore, the present definition of the level shift operator differs from that of \cite{MSB1} by a sign.} We do not present the explicit calculations, which are rather standard (albeit a bit lengthy), but give the results only. It suffices to consider $\Lambda_e$ with $e\geq 0$, i.e. with $e=e_j$ in \fer{45}.\footnote{It is not hard to see that $\Lambda_{-e}=-J_\s\Lambda_e J_\s$, where $J_\s$ is the modular conjugation associated to the von Neumann algebra ${\cal B}(\h_\s)\otimes\bbbone\subset {\cal B}(\h_\s\otimes\h_\s)$ and the trace state \fer{Strace}.}

{} For $x>0$ and $h\in L^2(\rx^3,\d^3k)$ we define
\begin{equation}
\sigma^\pm_h(x) = 2 \pi x^2 \frac{\e^{\pm\beta x}}{\sinh(\beta x)} \int_{S^2} |h(2x,\Sigma)|^2\d\Sigma.
\label{s10}
\end{equation}

$\bullet$  The eigenspace of $L_\s$ associated with $e=0$ is spanned by $\{\varPhi_1\otimes\varPhi_1,\varPhi_2\otimes\varPhi_2 ,\varPhi_3\otimes\varPhi_3, \varPhi_4\otimes\varPhi_4\}$, where the $\varPhi_j$ are given in \fer{Sbasis}. In this basis we have 
\begin{eqnarray}
\Lambda_0 &=& \i\left\{ \mu_1^2 \sigma^-_{g_1}(B_1) + \lambda_1^2\sigma_g^-(B_1)\right\}
\left[
\begin{array}{cccc}
\e^{2\beta B_1} & 0 & -\e^{2\beta B_1} & 0 \\
0 & \e^{2\beta B_1} & 0 & -\e^{2\beta B_1} \\
-1 & 0 & 1 & 0 \\
0 & -1 & 0 & 1 
\end{array}
\right]
\label{s5}\\
&& +\i\left\{ \mu_2^2 \sigma^-_{g_2}(B_2) + \lambda_2^2\sigma_g^-(B_2)\right\}
\left[
\begin{array}{cccc}
\e^{2\beta B_2} &  -\e^{2\beta B_2} & 0 & 0 \\
-1 & 1 & 0 & 0 \\
0 & 0 & \e^{2\beta B_2} & -\e^{2\beta B_2} \\
0 & 0 & -1 & 1 
\end{array}
\right]
\label{s6}
\end{eqnarray}
We see that $\Lambda_0$ is the sum of two terms \fer{s5} and \fer{s6}, representing the (independent) interaction of the reservoirs with spin 1 and spin 2, respectively. $\Lambda_0$ does not depend on the energy-conserving interaction (on $\varkappa_{1,2}$ and $\nu_{1,2}$). Indeed, those interactions leave the diagonal of the density matrix (in the energy basis) invariant. The contributions coming from the local ($\mu_{1,2}$) and from the collective ($\lambda_{1,2}$) couplings enter $\Lambda_0$ in the same way.

\medskip

$\bullet$ The eigenspace of $L_\s$ associated with $e=2B_1$ is spanned by $\{\varPhi_1\otimes\varPhi_3, \varPhi_2\otimes\varPhi_4 \}$, where the $\varPhi_j$ are given in \fer{Sbasis}. In this basis we have 
\begin{eqnarray}
\Lambda_{2B_1}
 &=& \left\{ \i[\mu_1^2\sigma_{g_1}^-(B_1)+\lambda_1^2\sigma_g^-(B_1)] \textstyle\frac{1+\e^{2\beta B_1}}{2} +\mu_1^2 r_{g_1}(B_1)+\lambda_1^2 r_g(B_1) \right\}
\left[
\begin{array}{cc}
1 & 0 \\
0 & 1
\end{array}
\right]
\nonumber\\
&&+\i\left\{ \mu_2^2 \sigma^-_{g_2}(B_2) + \lambda_2^2\sigma_g^-(B_2)\right\}
\left[
\begin{array}{cc}
\e^{2\beta B_2} &  -\e^{2\beta B_2} \\
-1 & 1\\
\end{array}
\right]\nonumber\\
&& 
+\i[ \kappa_1^2\sigma_f(0) +\nu^2_1\sigma_{f_1}(0)]
\left[
\begin{array}{cc}
1 & 0\\
0 & 1
\end{array}
\right]\nonumber\\
&&
-2\kappa_1\kappa_2 r
\left[
\begin{array}{cc}
1 & 0\\
0 & -1
\end{array}
\right],
\label{s7}
\end{eqnarray}
where  $r$ is given in \fer{pvr} and 
\begin{equation}
r_g(x) = \frac{1}{2} \ {\rm P.V.} \int_{\rx\times S^2} u^2 |g(|u|,\Sigma)|^2\coth\left(\textstyle\frac{\beta |u|}{2}\right) \frac{1}{u-2x}\ \d u\d\Sigma.
\label{s8}
\end{equation}

\medskip

$\bullet$ The eigenspace of $L_\s$ associated with $e=2B_2$ is spanned by $\{\varPhi_1\otimes\varPhi_2, \varPhi_3\otimes\varPhi_4 \}$, where the $\varPhi_j$ are given in \fer{Sbasis}. We obtain $\Lambda_{2B_2}$ in this basis by switching all indices $1\leftrightarrow 2$ labelling the spins in \fer{s7}.

\medskip

$\bullet$ The eigenspace of $L_\s$ associated with $e=2(B_2-B_1)$ is spanned by $\varPhi_3\otimes\varPhi_2$, where the $\varPhi_j$ are given in \fer{Sbasis}. The level shift operator $\Lambda_{2(B_2-B_1)}$ is just a number (times the projection operator $|\varPhi_3\otimes\varPhi_2\rangle\langle \varPhi_3\otimes\varPhi_2|$). We have
\begin{eqnarray}
\Lambda_{2(B_2-B_1)} &=& \i\Big[ \mu_1^2\sigma_{g_1}(B_1) +\lambda_1^2\sigma_g(B_1) + \mu_2^2\sigma_{g_2}(B_2) +\lambda_2^2\sigma_g(B_2)\nonumber\\
&& \qquad 
+(\kappa_1-\kappa_2)^2\sigma_f(0) +\nu_1^2\sigma_{f_1}(0) +\nu_2^2\sigma_{f_2}(0)  \Big]\nonumber\\
&&+\mu_1^2 r_{g_1}(B_1) + \lambda_1^2r_g(B_1) -\mu_2^2 r_{g_2}(B_2) - \lambda_2^2r_g(B_2).
\label{s9}
\end{eqnarray}

\medskip

$\bullet$ The eigenspace of $L_\s$ associated with $e=2(B_2+B_1)$ is spanned by $\varPhi_1\otimes\varPhi_4$, where the $\varPhi_j$ are given in \fer{Sbasis}. The level shift operator $\Lambda_{2(B_2+B_1)}$ is just a number (times the projection operator $|\varPhi_1\otimes\varPhi_4\rangle\langle \varPhi_1\otimes\varPhi_4|$). We have
\begin{eqnarray}
\Lambda_{2(B_1+B_2)} &=& \i\Big[ \mu_1^2\sigma_{g_1}(B_1) +\lambda_1^2\sigma_g(B_1) + \mu_2^2\sigma_{g_2}(B_2) +\lambda_2^2\sigma_g(B_2)\nonumber\\
&& \qquad 
+(\kappa_1+\kappa_2)^2\sigma_f(0) +\nu_1^2\sigma_{f_1}(0) +\nu_2^2\sigma_{f_2}(0)  \Big]\nonumber\\
&&-\mu_1^2 r_{g_1}(B_1) - \lambda_1^2r_g(B_1) -\mu_2^2 r_{g_2}(B_2) - \lambda_2^2r_g(B_2).
\label{s9'}
\end{eqnarray}

\subsection{Resonance energies and resonance vectors}
\label{rerv}

We give here the eigenvalues and eigenvectors of the level shift operators of the last section. To ease notation we assume from now on that the form factors governing the energy-exchange interactions with all reservoirs are the same, and that those governing the energy-conserving ones are too,
\begin{equation}
g_1=g_2=g,\qquad \mbox{and}\qquad f_1=f_2=f.
\label{s10'}
\end{equation}
In order to distinguish different contributions, we still keep all the coupling constans distinct. We present here the resonance data: the eigenvalues $\delta_e^{(s)}$ and the resonance states $\eta_e^{(s)}$ and $\widetilde\eta_e^{(s)}$, defined in \fer{19}. For notational convenience, we set
\begin{equation}
e_j = \e^{2\beta B_j},\ \ \ j=1,2.
\label{s11}
\end{equation}
The resonance vectors below are written in coordinates associated to the bases given in the above section.

$\bullet$ The resonance data associated to $e=0$ are
\begin{equation}
\begin{array}{lll}
\delta_0^{(1)}=0, 
& 
\eta_0^{(1)} = \left[
\begin{array}{c}
1\\1\\1\\1
\end{array}
\right],
&
\widetilde\eta_0^{(1)} = \frac{1}{{\rm Tr} \e^{-\beta H_\s}}
\left[
\begin{array}{c}
1/\sqrt{e_1e_2} \\  \sqrt{e_2/e_1}  \\ \sqrt{e_1/e_2} \\ \sqrt{e_1e_2}
\end{array}
\right]
\\ 
\delta_0^{(2)}=\i (\mu_2^2+\lambda_2^2) \sigma_g(B_2), 
& 
\eta_0^{(2)} = \left[
\begin{array}{c}
-e_2\\1\\-e_2\\1
\end{array}
\right],
&
\widetilde\eta_0^{(2)} = \frac{\sqrt{e_1/e_2} }{{\rm Tr}\e^{-\beta H_\s}} 
\left[
\begin{array}{c}
-1/e_1\\ 1/e_1 \\ -1 \\ 1
\end{array}
\right]
\\ 
\delta_0^{(3)}=\i(\mu_1^2+\lambda_1^2)\sigma_g(B_1), 
& 
\eta_0^{(3)} = \left[
\begin{array}{c}
-e_1\\-e_1\\ 1\\1
\end{array}
\right],
&
\widetilde\eta_0^{(3)} = \frac{\sqrt{e_2/e_1} }{{\rm Tr}\e^{-\beta H_\s}} 
\left[
\begin{array}{c}
-1/e_2\\ -1 \\ 1/e_2 \\ 1
\end{array}
\right]
\\ 
\delta_0^{(4)}=\delta_0^{(2)}+\delta_0^{(3)},
& 
\eta_0^{(4)} = \left[
\begin{array}{c}
e_1e_2\\-e_1\\ -e_2\\1
\end{array}
\right],
&
\widetilde\eta_0^{(4)} = \frac{1}{{\rm Tr}\e^{-\beta H_\s}} \frac{1}{\sqrt{e_1e_2}}
\left[
\begin{array}{c}
1\\ -1 \\ -1 \\ 1
\end{array}
\right]
\end{array}
\label{ss1}
\end{equation}

$\bullet$ The level shift operator \fer{s7} has the form
\begin{equation}
\Lambda_{2B_1} = A\bbbone +
\left[
\begin{array}{cc}
e_2B +C & -e_2B \\
-B & B-C
\end{array}
\right],
\label{s15}
\end{equation}
where $A=\i(\lambda_1^2+\mu_1^2)\textstyle\frac12\sigma_g(B_1)+\i (\kappa^2_1 +\nu^2_1)\sigma_f(0) -(\lambda_1^2+\mu_1^2) r_g(B_1)$,  $B = \i(\lambda_2^2+\mu_2^2) \sigma^-_g(B_2)$ and  $C=-2\kappa_1\kappa_2 r$ (recall definitions \fer{57}, \fer{58}, \fer{pvr} and \fer{s10}, \fer{s8}, \fer{s11}).

The resonance energies associated to $e=2B_1$ are
\begin{equation}
\delta_{2B_1}^{(\pm)} = A +\textstyle\frac12 B(1+e_2) \pm\textstyle\frac12\left[ B^2(1+e_2)^2 +4C[B(e_2-1)+C]\right]^{1/2},
\label{s16}
\end{equation}
where the square root is the principal branch (with branch cut on the negative real axis). The associated resonance eigenvectors are
\begin{equation}
\begin{array}{ll}
\eta_{2B_1}^{(\pm)} = 
\left[
\begin{array}{c}
1\\
y_\pm
\end{array}
\right], \quad
& 
\widetilde\eta_{2B_1}^{(\pm)} = \frac{1}{1+e_2(\bar y_\pm)^2}
\left[
\begin{array}{c}
1\\
e_2\bar y_\pm
\end{array}
\right],
\end{array}
\label{s17}
\end{equation}
where $\bar y_\pm$ is the complex conjugate of $y_\pm= 1+\textstyle\frac{A+C-\delta_{2B_1}^{(\pm)}}{e_2B}$.

{\it Remark.\ }If $C=0$ then the eigenvalues \fer{s16} reduce to $A+B(1+e_2)$ and $A$ and the resonance vectors have easy expressions too.

$\bullet$ The resonance data for $e=2B_2$ is obtained from that of $e=2B_1$ by switching indices $1\leftrightarrow 2$ labelling spin 1 and spin 2.

$\bullet$ The resoance energy associated to $e=2(B_2-B_1)$ is given by \fer{s9}, and we have $\eta_{2(B_2-B_1)}=\widetilde\eta_{2(B_2-B_1)} =\varPhi_3\otimes\varPhi_2$.

$\bullet$ The resoance energy associated to $e=2(B_2+B_1)$ is given by \fer{s9'}, and we have $\eta_{2(B_2+B_1)}=\widetilde\eta_{2(B_2+B_1)} = \varPhi_1\otimes\varPhi_4$.

\subsection{Proof of Theorem \ref{thmrates}}

In order to obtain the thermalization and decoherence rates, we simply need to calculate the imaginary parts of the second-order contributions to the resonance energies, calculated in the previous section, and invoke Theorem \ref{thmz1}. Relations \fer{52}, \fer{55} and \fer{56} are immediate from \fer{ss1}, \fer{s9} and \fer{s9'}. Relation \fer{53} follows from \fer{s16}. This completes the proof of the Theorem. \hfill $\blacksquare$

\section{Proof of Theorem \ref{finitet}}
 \label{sectthmfinitet}

{}For a density matrix diagonal in
the energy basis, $\rho={\rm diag}(p_1,p_2,p_3,p_4)$, we have 
$D(\rho) = -2\min\{\sqrt{p_1p_4}, \sqrt{p_2p_3}\}$ (see \ref{60}). By the property of return to equilibrium, the reduced
density matrix of $\s_1+\s_2$ approaches the Gibbs state modulo an
error,
\begin{equation}
\lim_{t\rightarrow\infty}\rho_t = \rho_{\s,\beta} +O(\varkappa^2),
\label{63}
\end{equation}
where, in the energy basis,
$$
\rho_{\s,\beta}=\frac{1}{{\rm Tr\,}\e^{-\beta H_\s}}\ {\rm diag}
(\e^{-\beta(B_1+B_2)},\e^{-\beta(B_1-B_2)},\e^{-\beta(-B_1+B_2)},\e^{-\beta(-B_1-B_2)}).
$$
We have $D(\rho_{\s,\beta})=-2\frac{1}{{\rm Tr\,}\e^{-\beta
H_\s}}$ and so \fer{63} implies
\begin{equation}
\lim_{t\rightarrow\infty}D(\rho_t) = -2\frac{1}{{\rm
Tr\,}\e^{-\beta H_\s}} +O(\varkappa^2). \label{64}
\end{equation}
The concurrence vanishes if $D(\rho_t)\leq 0$. By a Dyson series
expansion, one can show that the error term in \fer{64} is uniform
in $\beta$ for $\beta\leq \beta_0$, where $\beta_0<\infty$ is any
fixed number. (See also \cite{FM} for a more detailed analysis and
a better bound.) Therefore, if $\beta\leq\beta_0$, then there
exists a constant $C>0$ (depending only on $\beta_0$) s.t. if
${\rm Tr\,}\e^{-\beta H_\s}\leq C\varkappa^{-2}$ then the right
side of \fer{64} is strictly negative. Then the existence
of a finite $t_0$ follows from the continuity of $t\mapsto D(\rho_t)$.  Next suppose that return to equilibrium takes place at rate
$\gamma$, i.e., that $\|\rho_t-\rho_{\s,\beta}+O(\varkappa^2)\|\leq C\e^{-\gamma t}$. 
Then $\xi(\rho_t) = \xi(\rho_{\s,\beta}) + O(\varkappa^2)
+O(\e^{-\gamma t})$ and by perturbation theory $D(\rho_t) = -2\frac{1}{{\rm Tr\,}\e^{-\beta H_\s}}+ O(\varkappa^2)
+O(\e^{-\gamma t})$. 
Standard estimates on return to equilibrium show that the
remainder term $O(\e^{-\gamma t})$ is uniform in $T$ varying in
compacta in $(0,\infty)$ \cite{MSB1,MSB3,FM}. Thus there is a
constant $c'>0$ such that $D(\rho_t)\leq 0$ for $t\geq\gamma^{-1} \ln[c'\,{\rm Tr}\,\e^{-\beta H_\s}]$. \hfill $\blacksquare$

\section{Proof of Theorem \ref{thm3}}

The dynamics of the matrix
elements in \fer{m3} are obtained according to Theorem
\ref{thm01}. For instance, with $\Phi_{ij}=\Phi_i\otimes\Phi_j$ (\fer{Sbasis})
\begin{eqnarray}
p_1(t) &=& pA_t(11;11) +(1-p) A_t(11;44) \nonumber\\
 &=& \sum_{s=1}^4 \e^{\i t\varepsilon_0^{(s)}}\scalprod{\widetilde\eta_0^{(s)}}{\Phi_{11}}\left\{ p \scalprod{\Phi_{11}-\Phi_{44}}{\eta_0^{(s)}} +\scalprod{\Phi_{44}}{\eta_0^{(s)}}\right\},
\label{m1'}
\end{eqnarray}
where the resonance energies and resonance vectors are given in
Section \ref{rerv}. Since $\rho_t$ is a density matrix, the diagonal
elements
\begin{equation}
x_j(t):=[\rho_t]_{jj},
\label{m12}
\end{equation}
$j=1,2,3,4$, must be non-negative and add up to one. It follows
from \fer{m3} that
\begin{equation}
p_j(t) = x_j(t) +O(\varkappa^2),\qquad j=1,2,3,4,
\label{m13}
\end{equation}
and
\begin{equation}
\rho_t =
\left[
\begin{array}{cccc}
x_1 & 0 & 0 & \alpha\\
0 & x_2 & 0 & 0 \\
0 & 0 & x_3 & 0 \\
\overline\alpha & 0 & 0 & x_4
\end{array}
\right] +O(\varkappa^2).
\label{m14}
\end{equation}
We recall that the remainder term in the previous formula, as well
in all that follows, is uniform in $t\geq 0$.  It is sometimes
more practical to consider the $x_j$ instead of the $p_j$ since
the former are known to be non-negative. The spectrum of $\xi_t$,
\fer{62}, with $\rho$ replaced by $\rho_t$, \fer{m14}, is
\begin{equation}
{\rm spec}(\xi_t)=\left\{ x_2x_3, x_2x_3, [\sqrt{x_1x_4}\pm|\alpha|]^2\right\} +O(\varkappa^2).
\label{m15}
\end{equation}
Let $D= D(\rho_t)$ be the quantity defined in \fer{61}. In order
to calculate $D$, we need to know which of the eigenvalues of
$\xi$ is the largest one.

We have the following expressions for $x_1,\ldots, x_4$
(see \fer{m13} and e.g. \fer{m1'} for $x_1$)
\begin{eqnarray}
x_1 &=& \frac{\e^{-\beta(B_1+B_2)}}{{\rm Tr}\e^{-\beta H_\s}}
 \left[ (1-\e^{-t\delta_2})(1-\e^{-t\delta_3}) \right.+\nonumber\\
&& \left.+p\left\{\e^{-t\delta_2}(e_2+1)+\e^{-t\delta_3}(e_1+1)
 +\e^{-t\delta_4}(e_1e_2-1)\right\}\right] +O(\varkappa^2)\label{x1}\\
x_2 &=& \frac{\e^{-\beta(B_1-B_2)}}{{\rm Tr}\e^{-\beta H_\s}} \left[ (1-\e^{-t\delta_2}) e_2^{-1}\{p(e_2+1)-1\} +(1-\e^{-t\delta_3}) \{-p(e_1+1)+1\} \right.\nonumber \\
&&\left.+(1-\e^{-t\delta_4})e_2^{-1}\{p(e_1e_2-1)+1\}\right] +O(\varkappa^2)\label{x2} \\
x_3 &=& \frac{\e^{-\beta(-B_1+B_2)}}{{\rm Tr}\e^{-\beta H_\s}}\left[ (1-\e^{-t\delta_2}) \{-p(e_2+1)+ 1\} +(1-\e^{-t\delta_3}) e_1^{-1}\{p(e_1+1)-1\} \right.\nonumber \\
&&\left.+(1-\e^{-t\delta_4})e_1^{-1}\{p(e_1e_2-1)+1\}\right] +O(\varkappa^2)\label{x3} \\
x_4 &=& \frac{\e^{-\beta(-B_1-B_2)}}{{\rm Tr}\e^{-\beta H_\s}} \left[ (1-p)(1+e_2^{-1}\e^{-t\delta_2}) (1+e^{-1}_1\e^{-t\delta_3})\right. +\nonumber\\
&& \left. +p(1-\e^{-t\delta_2})(1-\e^{-t\delta_3}) \right]
+O(\varkappa^2).\label{x4}
\end{eqnarray}
In the above expressions, we have set for short
\begin{equation}
e_j:=\e^{2\beta B_j}\quad \mbox{and}\quad
\delta_s:=\delta_0^{(s)},\ s=2,3,4 \label{mm-1}
\end{equation}
(recall that $\delta_0^{(1)}=0$ and that
$\delta_4=\delta_2+\delta_3$, see after \fer{delta2}). We have also used
the mean value theorem to obtain the estimate
$$
\e^{\i t\varepsilon_0^{(s)}} =\e^{\i t(\delta_s + O(\varkappa^4))}
= \e^{\i t\delta_s} +O(\varkappa^2),\ \ s=2,3,4,
$$
which holds uniformly in $t\geq 0$, provided $\delta_s>0$.

The off-diagonal matrix element is estimated by
\begin{equation}
\alpha(t) = \e^{\i t\varepsilon_{2(B_1+B_2)}}\alpha(0) = \e^{\i
t\Re \varepsilon_{2(B_1+B_2)}} \e^{-t\delta_5}\alpha(0) +
O(\varkappa^2), \label{m11}
\end{equation}
where $\delta_5$ is given in \fer{deltas}, and where
$\alpha(0)$ is linked to $a_1, a_2$ by \fer{m5}.

The above expressions for $x_1,\ldots x_4$ can be used to arrive
at the following result.
\begin{lem}
\label{mlem1} (a) We have, uniformly in $t\geq 0$ and in $p\in [0,1]$,
\begin{equation}
x_1x_4-x_2x_3 = \e^{-t(\delta_2+\delta_3)}p(1-p)
+O({\varkappa^2)}. \label{m19}
\end{equation}

(b) We have, uniformly in $t\geq 0$,
\begin{equation}
x_1x_4\geq [{\rm Tr}\e^{-\beta H_\s}]^{-2} p(1-p) +O(\varkappa^2).
 \label{m41}
\end{equation}
\end{lem}

\noindent
 {\bf Proof of Lemma \ref{mlem1}.\ } Relation \fer{m19}
is obtained by direct calculation from \fer{x1}-\fer{x4}. To show
(b), we note that by \fer{x1}, and since $1\geq p$,
$$
x_1\geq \frac{\e^{-\beta(B_1+B_2)}}{{\rm Tr}\e^{-\beta H_\s}} p\
f(t) +O(\varkappa^2),
$$
with $f(t)=1+\e^{-t\delta_2}e_2 +\e^{-t\delta_3}e_1 +\e^{-t\delta_4}e_1e_2\geq 1$.
Moreover, by \fer{x4}, 
$$
x_4\geq \frac{\e^{-\beta(-B_1-B_2)}}{{\rm Tr}\e^{-\beta H_\s}} (1-p) +O(\varkappa^2),
$$
uniformly in $t\geq
0$. This ends the proof of Lemma \ref{mlem1}.\hfill $\blacksquare$

\bigskip
In what follows we distinguish two cases (inequalities A and B
below). We denote by $C$ generic constants which are independent
of $p$ and $\varkappa$, but whose values may change from
expression to expression.

\begin{itemize}
\item[A] $x_2x_3\geq [\sqrt{x_1x_4}+|\alpha|]^2
+O(\varkappa^2)$.\\
We first prove that if $t\geq t_A$ (see \fer{m17}) then inequality A holds. It follows from \fer{m19} that
inequality A holds if $\e^{-t(\delta_2+\delta_3)} p(1-p)
+2|\alpha| \sqrt{x_1x_4} +|\alpha|^2\leq C\varkappa^2$, for some
$C>0$. Since $x_j\leq 1+O(\varkappa^2)$ and $|\alpha|\leq
\e^{-t\delta_5}\sqrt{p(1-p)} +O(\varkappa^2)$ (see also \fer{m11}
and \fer{star}), this condition is satisfied provided
\begin{equation}
\e^{-t\delta_5}\sqrt{p(1-p)}\leq C\varkappa^2 \qquad
\mbox{and}\qquad \e^{-t(\delta_2+\delta_3)}p(1-p)\leq
C\varkappa^2. \label{m20}
\end{equation}
Conditions \fer{m20} hold if $t\geq t_A$, see \fer{m17}. Note
that only the first two terms in the max on the r.h.s. of
\fer{m17} are  needed for this argument, the last one will be used
below.

Next we show that
\begin{equation}
D=-2\max\{\sqrt{x_1x_4},|\alpha|\}+O(\varkappa^2). \label{m40}
\end{equation}
Inequality A implies that the largest
eigenvalue of $\xi_t$ is $x_2x_3+O(\varkappa^2)$, see \fer{m15}.
To calculate $D$, \fer{61}, we need to take the square roots of the
eigenvalues of $\xi$. Using \fer{m41} together with \fer{m19} we
obtain
\begin{equation}
x_2x_3\geq  Cp(1-p) -\e^{-t(\delta_2+\delta_3)} p(1-p)
+O(\varkappa^2),
 \label{m44}
\end{equation}
and consequently, since $t\geq t_A$ and hence $t\geq
C\frac{1}{\delta_2+\delta_3}$, we have
\begin{equation}
x_2x_3\geq C p(1-p) +O(\varkappa^2).
 \label{m45}
\end{equation}

We conclude that for $\frac{\varkappa^2}{p(1-p)}$ small enough
($|\varkappa|\leq \varkappa_0\sqrt{p(1-p)}$ for some $\varkappa_0$
independent of $p$), we have  the following expressions for the
square roots of the eigenvalues of $\xi_t$:
\begin{eqnarray}
\sqrt{(\sqrt{x_1x_4}\pm |\alpha|)^2+O(\varkappa^2)} &=& \big|
\sqrt{x_1x_4}\pm |\alpha|\big|  +O(\varkappa^2) \label{m43}\\
\sqrt{x_2x_3+O(\varkappa^2)} &=& \sqrt{x_2x_3}+O(\varkappa^2).
 \label{m44.1}
\end{eqnarray}
Using expressions \fer{m43} and \fer{m44.1} in \fer{m15} we arrive
at \fer{m40}.

We are now ready to complete the proof of point A of Theorem \ref{thm3}. We have $|\alpha|\leq C\varkappa^2$, see \fer{m20}, and hence
$$
 \max\{ \sqrt{x_1x_4},
|\alpha|\} \geq C\sqrt{p(1-p)}.
$$
Therefore, by \fer{m40}, $D<0$
and by \fer{60}, the concurrence vanishes.

\item[B] $x_2x_3\leq [\sqrt{x_1x_4}+|\alpha|]^2
+O(\varkappa^2)$.\\
Due to \fer{m19} and \fer{m41}, inequality B holds if
$$
\e^{-t(\delta_2+\delta_3)}p(1-p) +C\e^{-t\delta_5} p(1-p)
+\e^{-2t\delta_5} p(1-p) \geq \widetilde C\varkappa^2,
$$
for some $\widetilde C>0$. The latter condition is satisfied if
either of the tree positive summands on the left hand side are
bounded below by $\widetilde C\varkappa^2$, i.e. if
\begin{eqnarray}
t&\leq& \max\left\{ \frac{1}{\delta_5}\ln \left( C_B
\frac{p(1-p)}{\varkappa^2} \right), \frac{1}{\delta_2+\delta_3}
\ln\left(C_B \frac{p(1-p)}{ \varkappa^2}\right) \right\} \nonumber\\
&=& \frac{1}{\delta_2+\delta_3} \ln\left(C_B \frac{p(1-p)}{
\varkappa^2}\right).
\label{-100}
\end{eqnarray}
We have used here that $\delta_5\geq \delta_2+\delta_3$, see \fer{deltas}.

Next we analyze $D$, \fer{61}. The largest eigenvalue of $\xi_t$,
\fer{m15}, is $\nu_1= [\sqrt{x_1x_4}+|\alpha|]^2 +O(\varkappa^2)$.
Its square root is  $\sqrt\nu_1=\sqrt{x_1x_4}+|\alpha|
+O(\varkappa^2)$ (this follows from \fer{m41} and $|\varkappa|\leq
\varkappa_0\sqrt{p(1-p)}$). The quantity $D$ is obtained by
subtracting from $\sqrt\nu_1$ the terms
$\sqrt{(\sqrt{x_1x_4}-|\alpha|)^2+O(\varkappa^2)}$ and twice
$\sqrt{x_2x_3+O(\varkappa^2)}$. We are now showing that for $t\leq
t_B$, we have
\begin{eqnarray}
\sqrt{x_2x_3 +O(\varkappa^2)} &\leq & \frac{1}{10}
\sqrt\nu_1\qquad \mbox{and} \label{100} \\
\sqrt{(\sqrt{x_1x_4}-|\alpha|)^2+O(\varkappa^2)} &\leq&
\frac{1}{10} \sqrt\nu_1.\label{101}
\end{eqnarray}
It then follows that for $t\leq t_B$, we have $C(\rho_t)\geq
\frac{7}{10}\sqrt{\nu_1}$, and due to \fer{m41} the statement of point B in Theorem \ref{thm3} holds. \\
 It remains to show
\fer{100} and \fer{101}. We obtain the upper bound $x_2x_3\leq C(1-\e^{-t\delta_4})^2 +O(\varkappa^2)$ directly from expressions \fer{x2}, \fer{x3}. By taking this into account, together with \fer{m19}, \fer{m41} and $|\varkappa|\leq \varkappa_0\sqrt{p(1-p)}$, we see that (the square of) \fer{100} holds provided that $1-\e^{-t\delta_4} \leq C p(1-p)$ (where $C$ is small), which in turn is implied by
\begin{equation}
t\leq \frac{1}{\delta_4}\ln[Cp(1-p)+1].
\label{104}
\end{equation}
To summarize, condition \fer{104} implies \fer{100}. 

Our next task is to prove \fer{101}. By squaring this inequality we see that it is satisfied provided that $(\sqrt{x_1x_4}-|\alpha|)^2 \leq  \frac{1}{100} x_1x_4 +O(\varkappa^2)$. Invoking \fer{m19} and $|\varkappa|\leq \varkappa_0\sqrt{p(1-p)}$, we see that the last inequality holds provided
\begin{equation}
(\sqrt{x_1x_4}-|\alpha|)^2 \leq  Cp(1-p).
\label{105}
\end{equation}
Note that the l.h.s vanishes at $t=0$, so the inequality holds for small enough times. Let us set
\begin{equation}
\delta_-:=\min\{\delta_2,\delta_3\} \qquad\mbox{and}\qquad \delta_+:=\max\{\delta_2,\delta_3\}.
\label{106}
\end{equation}
It follows from \fer{x1} and \fer{x4} that
\begin{eqnarray}
x_1x_4&\leq& [{\rm Tr}\e^{-\beta H_\s}]^{-2}\left\{ p(1-p) \e^{-t\delta_-}(e_1+e_2+e_1e_2+1)\right.\nonumber \\
&&\left. \times(1+\e^{-t\delta_2}/e_1)(1+\e^{-t\delta_3}/e_1)  +C(1-\e^{-t\delta_+})^2\right\} +O(\varkappa^2).\qquad
\label{107}
\end{eqnarray}
We estimate
\begin{equation}
(1+\e^{-t\delta_2}/e_1)(1+\e^{-t\delta_3}/e_1)\leq (1+1/e_1)(1+1/e_2) +C(1-\e^{-t\delta_+}).
\label{108}
\end{equation}
Combining \fer{107} and \fer{108}, and using the definition \fer{mm-1} of $e_j$, we obtain the upper bound
\begin{equation}
x_1x_4\leq p(1-p)\e^{-t\delta_-} +C(1-e^{-t\delta_+})+O(\varkappa^2).
\label{109}
\end{equation}
Furthermore, if $t\leq \frac{1}{\delta_+}\ln[1/(1-C\varkappa^2)]$ for some constant $C>0$, then $C(1-e^{-t\delta_+})=O(\varkappa^2)$. The last upper bound on $t$ is implied by
\begin{equation}
t\leq \frac{1}{\delta_+}\ln[1 +C \varkappa^2].
\label{109.5}
\end{equation}
 Next we have $|\alpha|=e^{-t\delta_5}\sqrt{p(1-p)}+O(\varkappa^2)$, so we obtain for $t$ satisfying \fer{109.5}
\begin{equation}
\sqrt{x_1x_4}-|\alpha|\leq \sqrt{p(1-p)}\left[\e^{-\frac{t}{2}\delta_-}-\e^{-t\delta_5}\right]+O(\varkappa^2).
\label{110}
\end{equation}
We also know that $-C\varkappa^2\leq \sqrt{x_1x_4}-|\alpha|$, a fact that follows simply from the positivity of $\rho_t$ (see also \fer{m14}). Therefore,
\begin{equation}
(\sqrt{x_1x_4}-|\alpha|)^2\leq \max\left\{ C\varkappa^4, \ p(1-p)\left[\e^{-\frac{t}{2}\delta_-}-\e^{-t\delta_5}\right]^2+O(\varkappa^2) \right\}.
\label{111}
\end{equation}
We use this upper bound to see that \fer{105} holds provided
\begin{equation}
p(1-p)\left[\e^{-\frac{t}{2}\delta_-}-\e^{-t\delta_5}\right]^2 \leq Cp(1-p) +O(\varkappa^2).
\label{112}
\end{equation}
Since $|\varkappa|\leq \varkappa_0\sqrt{p(1-p)}$ inequality \fer{112} is implied by the bound $\e^{-\frac{t}{2}\delta_-}-\e^{-t\delta_5}\leq C$. We have $\e^{-\frac{t}{2}\delta_-}-\e^{-t\delta_5}=\e^{-\frac t2 \delta_-}[1-\e^{-t(\delta_5-\delta_-/2)}]\leq 1-\e^{-t(\delta_5-\delta_-/2)} $, so \fer{111} holds if $1-\e^{-t(\delta_5-\delta_-/2)} \leq C$, which in turn is implied by
\begin{equation}
t\leq \frac{C}{\delta_5-\delta_-/2}.
\label{113}
\end{equation}
We have thus shown that if $t$ satisfies \fer{-100}, \fer{104}, \fer{109.5} and \fer{113}, then the bounds \fer{100} and \fer{101} hold. Condition \fer{-100} is implied by condition \fer{104} since $|\varkappa|\leq \varkappa_0\sqrt{p(1-p)}$. Thus all above constraints on $t$ are verified for $t\leq t_B$, see \fer{m18}. This shows point B of Theorem \ref{thm3}.
\end{itemize}
The proof of Theorem \ref{thm3} is complete.\hfill $\blacksquare$

\section{Davies generator and level shift operators}
\label{dglso}

We take all the coupling constants in \fer{4}-\fer{7} to be proportional to some $\varkappa$. Let $\alpha_\varkappa^t$ denote the reduced Schr\"odinger dynamics of the qubits.
\begin{prop}
\label{propos}
Let $\rho$ be any density matrix of the qubits. Then 
\begin{equation}
\lim_{\varkappa\rightarrow 0}\sup_{\tau\geq 0}\|\alpha_0^{-\tau/\varkappa^2}\circ\alpha_\varkappa^{\tau/\varkappa^2}(\rho) -\e^{\tau K^\sharp}\rho\|=0,
\label{u2}
\end{equation}
where the operator $K^\sharp:l^1(\h_\s)\rightarrow l^1(\h_\s)$ maps density matrices into density matrices. Viewing $l^1(\h_\s)$ as $\h_\s\otimes\h_\s$, $K^\sharp$ is an operator on the latter tensor product. As such, it leaves spectral subspaces of $L_\s$, \fer{16'} invariant. On the subspace ${\rm Ran}P_{L_\s=e}$, $K^\sharp$ acts as $(\i\Lambda_e)^*$ (adjoint of level shift operator \fer{lso}).
\end{prop}

{\it Remarks.\ } {\bf (1)} Relation \fer{u2} implies that $\e^{\tau K^\sharp}$ maps density matrices to density matrices: indeed, $\alpha_0^{-\tau/\varkappa^2}\circ\alpha_\varkappa^{\tau/\varkappa^2}(\rho)$ is a density matrix. $K^\sharp$ is thus the Davies generator.

{\bf (2)} As defined in \fer{lso}, $\Lambda_e$ is proportional to $\varkappa^2$, but in the proposition, we prefer to reinterpret the level shift operator to be independent of $\varkappa$ (divide \fer{lso} by $\varkappa^2$).

\medskip

{\it Proof of Proposition \fer{propos}.\ } Our proof is partly inspired by \cite{DJ}. We have 
\begin{eqnarray}
[\alpha_0^{-t}\circ\alpha_\varkappa^{t}(\rho)]_{mn} &=&\scalprod{\Psi_0}{\e^{\i tL_\varkappa} \e^{-\i tL_0}(|\Phi_n\rangle\langle\Phi_m|\otimes\bbbone_\r) \e^{\i tL_0}\e^{-\i tL_\varkappa}\Psi_0}\nonumber\\
&=&\e^{-\i t E_{mn}}\sum_{(k,l)\in{\cal C}(E_{nm})} A_t(m,n;k,l)[\rho]_{kl} +O(\varkappa^2).
\label{u01}
\end{eqnarray}
Here, $\Psi_0$ is the initial state represented in the GNS space and $L_\varkappa$ is the standard Liouville operator implementing the dynamics. We use that $L_0=L_\s+L_\r$, $L_\s=H_\s\otimes\bbbone-\bbbone\otimes H_\s$, $H_\s\Phi_n =E_n\Phi_n$, and Theorem \ref{thm01}. We denote $E_{mn}=E_m-E_n$ and, below, $\Phi_{mn}=\Phi_m\otimes\Phi_n$. Setting $t=\tau/\varkappa^2$, using the explicit form of $A_t$, \fer{35}, and the fact that the remainder in \fer{u01} is uniform in $t$, we obtain \fer{u2} with
\begin{eqnarray}
[\e^{\tau K^\sharp}\rho]_{mn} &=& \sum_{(k,l)\in{\cal C}(E_{mn})} A_\tau^\sharp(m,n;k,l)\, [\rho]_{kl} \label{u3}\\
A_\tau^\sharp(m,n;k,l) &=& \scalprod{\Phi_{mn}}{\e^{-\i\tau(\Lambda_{E_{mn}})^*}\Phi_{kl}}. \label{u4}
\end{eqnarray}
To arrive at \fer{u4}, we make use of the diagonalization formula 
$$
\sum_{s=1}^{{\rm mult}(E_{nm})} \e^{\i\tau\delta_{E_{nm}}^{(s)}}|\eta_{E_{nm}}^{(s)}\rangle\langle \widetilde\eta_{E_{nm}}^{(s)}| = \e^{\i\tau \Lambda_{E_{nm}}},
$$
implying (see \fer{u01} and \fer{35})
\begin{eqnarray}
A_\tau^\sharp(m,n;k,l) &=&\scalprod{\Phi_{lk}}{\e^{\i\tau\Lambda_{E_{nm}}}\Phi_{nm}}\nonumber\\
 &=&\overline{\scalprod{\Phi_{kl}}{J_\s\e^{\i\tau\Lambda_{E_{nm}}}J_\s\Phi_{mn}}}\nonumber\\
&=&\scalprod{\Phi_{mn}}{J_\s\e^{-\i\tau(\Lambda_{E_{nm}})^*}J_\s\Phi_{kl}},
\label{u5}
\end{eqnarray}
where $J_\s$ is the modular conjugation associated to the pair $({\frak M}_\s,\Omega_\s)$. Finally, one sees readily (see also \cite{MLSO}) that $J_\s\Lambda_eJ_\s=-\Lambda_{-e}$, so that \fer{u4} follows from \fer{u5}.

As an operator on $l^1(\h_\s)$, $K^\sharp$ has matrix elements $K_{mn,kl}$ defined by $K^\sharp |\Phi_k\rangle\langle\Phi_l| = \sum_{m,n}K_{mn,kl} |\Phi_m\rangle\langle\Phi_n|$, and by applying $\partial_\tau|_{\tau=0}$ to \fer{u3}, \fer{u4}, one obtains
$$
K_{mn,kl} = \scalprod{\Phi_m}{(K^\sharp |\Phi_k\rangle\langle\Phi_l|) \Phi_n} =
\left\{
\begin{array}{cl}
\scalprod{\Phi_{mn}}{(\i \Lambda_{E_{mn}})^*\Phi_{kl}} & \mbox{ if $(k,l)\in{\cal C}(E_{mn})$}\\
0 &  \mbox{ if $(k,l)\notin{\cal C}(E_{mn})$}
\end{array}
\right.
$$
This completes the proof of Proposition \fer{propos}.\hfill $\blacksquare$

\end{document}